\documentclass[11pt,a4paper]{article}

\usepackage[normalem]{ulem}
\usepackage[utf8]{inputenc}
\usepackage{amsfonts}
\usepackage{amsmath}
\usepackage{amsmath}
\usepackage{amssymb, amsmath, amsbsy}
\usepackage{amssymb}
\usepackage{authblk}
\usepackage{bigints}
\usepackage{color}
\usepackage{colortbl}
\usepackage{enumerate}
\usepackage{float}
\usepackage{geometry}
\usepackage{graphicx}
\usepackage{lineno}
\usepackage{lscape}
\usepackage{mathdots}
\usepackage{moreverb}
\usepackage{multirow}
\usepackage{setspace}
\usepackage{subfigure}

\title{Stability and decay of composite kinks/$Q$-balls solutions in a deformed $O(2N+1)$ linear sigma model}

\author[1,2]{A. Alonso-Izquierdo}
\author[2]{ D. Canillas Martínez}
\author[2]{C. Garzón Sánchez}
\author[1,2]{M.A. González León}
\author[1,3,4]{A. Wereszczynski}

\affil[1]{ Departamento de Matemática Aplicada, University of Salamanca,
Casas del Parque 2, 37008 - Salamanca, Spain.}
\affil[2]{ IUFFyM, University of Salamanca,
Plaza de la Merced 1, 37008 - Salamanca, Spain}
\affil[3]{ Institute of Theoretical Physics, Jagiellonian University, Lojasiewicza 11, Kraków, Poland}
\affil[4]{ International Institute for Sustainability with Knotted Chiral Meta Matter (WPI-SKCM2), Hiroshima University, Higashi-Hiroshima, Hiroshima 739-8526, Japan}

\date{}

\begin{document}

\maketitle

\begin{abstract}
   The defect-type solutions of a deformed $O(2N+1)$ linear sigma model with a real and $N$ complex fields in $(1+1)$-dimensional Minkowski spacetime are studied. All the solutions are analytically found for the $N=2$ case. Two types of solitons have been determined: (a) Simple solutions formed by a topological kink with or without the presence of a $Q$-ball. (b) Composite solutions. They are constituted by some one-parameter families of solutions which can be understood as a non-linear combination of simple solutions. The properties of all of those solutions and the analysis of their linear stability, as well as decay channels, are discussed.
\end{abstract}


\section{Introduction}

Kinks and $Q$-balls are particle-like solutions which arise in scalar field theories with the shared property that they retain their profile throughout temporal evolution. However, the reason that supports the stability of these solutions is different in each case. Kinks are static topological solutions \cite{manton2004topological,shnir_2018} whose stability is guaranteed by the topology of the configuration space. On the other hand, $Q$-balls  \cite{friedberg1976class, coleman1985q, coleman1986errata} are {\it stationary}, time-dependent, non-topological solutions whose stability is justified by the presence of a conserved Noether charge associated with the internal $U(1)$ symmetry group. In this scenario, some sufficient conditions that guarantee the stability of these objects have been established \cite{LEE1992,Dine2003,tsumagari2008some}.   Importantly, the time dependence of these solutions permits them to avoid the severe restrictions of Derrick's theorem \cite{derrick1964comments,manton2004topological}, allowing the existence of $Q$-balls in theories where usual topological solitons can not exist.

Both kinks and $Q$-balls have inherent physical interest. The topological nature of the kink solutions makes them an incredibly powerful tool to describe different phenomena in nature, where non-linearity is involved, such as optics \cite{Optics,Mollenauer2006,Schneider2004,Agrawall1995}, biochemistry \cite{biochemistry,Yakushevich2004}, cosmology \cite{cosmology,Kolb1990,Kibble1976,Vilenkin1994,Vachaspati2006} and physics of materials \cite{graphene, Bishop1980,Eschenfelder1981,Jona1993,Strukov}. On the other hand, $Q$-balls appear in diverse scenarios of the evolution of the early universe. Perhaps, the most compelling physical interest comes from the fact that they may be produced in the early stages of the universe in supersymmetric extensions of the Standard Model, where leptonic and barionic $Q$-balls can be found and Noether charge is related to the conservation of lepton and baryon numbers \cite{KUSENKO1997108}. It was argued that these solutions may play an important role in the baryogenesis by means of the Affleck-Dine mechanism \cite{Dine2003,AFFLECK1985361}. They were also considered a plausible candidate for Dark Matter \cite{KUSENKO199846}. Coupled to gravity they give rise to boson stars, which are hypothetical compact self-gravitating objects \cite{Wheeler:1955zz, Kaup:1968zz, Ruffini:1969qy}.

Of course, due to their time-dependence, $Q$-balls are rather mathematically complicated solutions with a little chance for analytical results, although $Q$-balls in exact form are also known \cite{bazeia2016exact,Alonso2023_1}. Therefore, despite of the significant progress in the understanding of their properties, many fundamental problems still await for a comprehensive explanation. This, for example, concerns the stability of $Q$-balls. The issue was thoroughly analyzed in the seminal papers by Kolokolov \cite{kolokolov1973stability} and by Friedberg, Lee and Sirling \cite{friedberg1976class} leading to a sufficient and necessary condition determining the classical stability of $Q$-balls. This depends on the number of eigenvalues of the second order small fluctuation operator valued on the $Q$-ball. According to the criterion, if this operator has only one negative eigenvalue and the derivative of the Noether charge with respect to the internal frequency $\omega$ is negative, the classical stability of this non-topological soliton is guaranteed. Otherwise, the Q-balls are unstable.
It was thought that the presence of $Q$-balls always implies the existence of at least one negative eigenvalue and, therefore, the decrease of the Noether charge as a function of the internal frequency is a necessary condition. However, a counterexample of this result has recently been found \cite{Alonso2023_2}. When the Q-ball is interwoven with a kink, the spectrum of the fluctuation operator of this solution may consists of only positive eigenvalues, which stabilizes the solution. Hence, the nonlinear interaction between the complex and real field can act as a stabilizing factor for both $Q$-balls and lumps or kinks.
Likewise, the model addressed by Friedberg, Lee and Sirling the previously mentioned model involves a complex scalar field coupled with one real field, where the terms depending on the complex field exhibits an internal $U(1)$-symmetry which accommodates the $Q$-ball solutions.

In the present paper we shall investigate a deformed $O(2N+1)$ non-linear Sigma model in $(1+1)$ space-time dimensions (which generalizes the aforementioned model) where a real scalar field is coupled with $N$ complex fields, each of them having an independent internal $U(1)$-symmetry. As a consequence, $N$ Noether charges $Q_i$ emerge in this system and also every single $Q$-ball (carrying only one non zero charge $Q_i$) can give rise to composite defects in conjunction with the kink arising in the real field sector of the theory. Moreover, there exists the possibility of finding composite defects consisting of a kink and several $Q$-balls, forming a solution which carries several non-vanishing Noether charges $Q_i$.

The aim of the present paper is to study the existence of this type of solutions and the stability, trying to understand the possible $Q$-ball stabilization channels due to the presence of the kink. Furthermore we investigate dynamics of the perturbed composite $Q$-ball solutions identifying possible decay channels. This may be viewed as the first step in a more extensive and systematic analysis of the dynamics of $Q$-balls (for recent results on dynamics of $Q$-balls see e.g., \cite{Xie:2021glp, Saffin:2022tub}).

The organization of the article is as follows: in Section \ref{Section:2} we introduce the deformed $O(2N+1)$ non-linear sigma model addressed in this paper and describe its basic properties. For the sake of concreteness, we shall focus on the deformed $O(5)$ model, i.e., where two different Noether charges arise.  In Section \ref{Section:3} we discuss the structure of the solutions in this model, including basic defects consisting of a single kink and two types of solutions where a topological kink and a $Q$-ball are attached; three families whose members are composed by two basic defects separated by an arbitrary distance; and finally one family combining three of these basic defects. The analytical expressions of all these solutions are obtained in this case. The stability and the different decay channels are also discussed. Finally, in Section \ref{Section:4} the conclusions of this work are exposed.

\section{The model}\label{Section:2}

In this paper we deal with a deformed $O(2N+1)$ non-linear sigma model immersed in a $(1+1)$-dimensional Minkowski spacetime, which involves the coupling between one real and $N$ complex scalar fields. The dynamics of this model is characterized by the action
functional
\begin{equation}\label{eq:Action_Model}
	S=\int d^2 x \Big[\dfrac{1}{2} \partial_\mu \phi \ \partial^\mu \phi + \dfrac{1}{2} \sum_{j=1}^{N}\partial_\mu \overline\psi_j \partial^\mu \psi_j- U\left(\phi,|\psi_1| , ..., |\psi_N| \right)\Big] \ ,
\end{equation}
where $\phi$ and $\psi_{j}$ with $j=1,\dots,N$ are respectively a real and $N$ complex scalar fields. That is, $\phi\in\text{Maps}\left(\mathbb{R}^{1,1},\mathbb{R}\right)$ and $\psi_{j} \in \text{Maps}\left(\mathbb{R}^{1,1},\mathbb{C}\right)$. In (\ref{eq:Action_Model}), $\overline \psi_j$ stands for the complex conjugate to $\psi_j$. As usually in this context, the Minkowski metric is chosen in the form $\eta_{\mu\nu}=\text{diag}\left(1,-1\right)$ and Einstein summation convention is only applied to space-time indexes. The potential term $U\left(\phi, |\psi_1|, ..., |\psi_N|\right)$ considered in (\ref{eq:Action_Model}) is given by the positive semi-definite expression,
\begin{equation}\label{eq:Potential_Model}
	U(\phi,|\psi_1| , ..., |\psi_N|)=\dfrac{1}{2}\Big(\phi^2+\sum_{j=1}^{N}|\psi_j|^2-1\Big)^2+\dfrac{1}{2}\sum_{j=1}^{N} \sigma_j^2|\psi_j|^2 \ ,
\end{equation}
with positive real parameters  $\sigma_j \in \mathbb{R}$, $\sigma_i\neq \sigma_j$, $\forall i\neq j$. We assume, without loss of generality:  $0<\sigma_1^2<\sigma_2^2<\ldots<\sigma_n^2$. These parameters give us a measure of the asymmetry of the model with respect to the rotationally invariant situation in the $O(2N+1)$ non-linear sigma model. The formula (\ref{eq:Potential_Model}) is a quartic polynomial in the real field $\phi$ and the modulus of the complex fields $\psi_i$.

The set of zeroes of $U$, i.e. the vacua manifold of the model, is given in this case by:
\begin{equation}
 \mathcal{M}=\{ v \in (\phi,\psi_{1},...,\psi_{N}) | U(v)=0 \}= \{ v_+, v_- \}
\end{equation}
with: $v_\pm=(\pm 1, 0,...,0) $. These solutions correspond to the global minima of the potential $U(\phi,|\psi_j| ; \sigma_j)$. In order to analyze the stability of the vacua, the Hessian matrix is evaluated at these solutions,
\begin{equation} \label{eq:vacum_hessian}
	\mathcal{H}\left[v_{ \pm}\right]=\left(\begin{smallmatrix}
		\dfrac{\partial^2 U}{\partial \phi^2} & ... & \dfrac{\partial^2 U}{\partial \phi  \, \partial |\psi_{N}|}\\
		\vdots  & \ddots  & \vdots\\
		\dfrac{\partial^2 U}{\partial \phi \, \partial |\psi_{N}|}& \dots & \dfrac{\partial^2 U}{\partial |\psi_{N}|^2}
	\end{smallmatrix}\right)
	= \text{diag} \left( 4, \sigma_1^2, \dots, \sigma_N^2 \right) \ .
 \end{equation}
As we can see, the diagonal entries are positive, therefore the vacua are stable solutions of the field equations, as expected. This condition is necessary for the existence of other finite energy solutions. The field equations derived from minimizing (\ref{eq:Action_Model}) read
\begin{equation}\label{eq:FieldEq_Model}
	\dfrac{\partial^2 \phi}{\partial t^2}-\dfrac{\partial^2 \phi}{\partial x^2}+\dfrac{\partial U}{\partial \phi}=0 \quad , \quad \dfrac{\partial^2 \psi_j}{\partial t^2}-\dfrac{\partial^2 \psi_j}{\partial x^2}+\dfrac{\psi_j}{|\psi_j|} \dfrac{\partial U}{\partial|\psi_j|}=0 \ .
\end{equation}
By construction, the model is invariant under the $U(1)\times \dots \times U(1)$ symmetry group transformations, $\psi_j \rightarrow e^{i \theta_j} \psi_j$, in each $j=1,\dots, N$. This symmetry leads to the existence of the $N$ conserved Noether charges
\begin{equation}
	Q_j=\dfrac{1}{2 i}\int_{-\infty}^\infty  dx \left(\overline\psi_j\partial_t \psi_j-\psi_j \partial_t\overline\psi_j\right) \ .
\end{equation}
As previously mentioned, we are interested in investigating the presence of defects formed by the coupling between a kink and $Q$-balls. For this reason the ansatz
\begin{equation}\label{eq:ansatz}
	\phi(x,t)=f(x) \quad , \quad \psi_j(x,t)=g_j(x)e^{i \omega_j t} \quad, \quad j=1,\dots,N
\end{equation}
will be employed. $f(x)$ and $g_j(x)$ are real functions and $\omega_j$ are the internal rotation frequencies associated to every complex field in the internal space. Substituting this ansatz in the functional energy $E$, it can be written in the form
\begin{equation} \label{eq:Energy}
	E[f, g_1,..., g_N]=\int_{-\infty}^\infty d x\Big[ \dfrac{1}{2}\left(\dfrac{d f}{d x}\right)^2+\dfrac{1}{2}\sum_{j=1}^{N}\left(\dfrac{d g_j}{d x}\right)^2+\dfrac{1}{2}\sum_{j=1}^{N} \omega_j^2 g_j^2+U(f,g_1,..., g_N)\Big]  \ ,
\end{equation}
where the potential in (\ref{eq:Energy}) is given by,
\begin{equation}\label{eq:Potential_Model_ansatz}
	U(f, g_1,..., g_N)=\dfrac{1}{2}\Big(f^2+\sum_{j=1}^{N}g_j^2-1\Big)^2+\dfrac{1}{2}\sum_{j=1}^{N} \sigma_j^2g_j^2 \ ,
\end{equation}
We are only interested in finite energy solutions. Therefore, the following asymptotic conditions have to be satisfied
\begin{equation}
    \lim _{x \rightarrow \pm \infty} f(x) = \pm 1 \ , \quad
    \lim _{x \rightarrow \pm \infty} \dfrac{d f}{d x}=0 \ , \quad
    \lim _{x \rightarrow \pm \infty} g_i(x)=\lim _{x \rightarrow \pm \infty} \dfrac{d g_i}{d x}=0 \ .
\end{equation}
Finally, their conserved Noether charges reduce with these assumptions to:
\begin{equation} \label{eq:Noether_charges}
	Q_j=\omega_j \int_{-\infty}^\infty dx \ g_j(x)^2 \ ,
\end{equation}
which we can grouped as the multi-component magnitude
\[
Q= (Q_1,Q_2, \dots, Q_n)  \ .
\]
Inserting ansatz (\ref{eq:ansatz}) into the field equations, (\ref{eq:FieldEq_Model}) can be reduced to the following system of ordinary differential equations
\begin{equation}\label{odes}
\dfrac{d^2 f}{d x^2}=\frac{\partial U}{\partial f}\quad , \quad \frac{d^2 g_j}{d x^2}=\dfrac{\partial U}{\partial g_j}-\omega_j^2 g_j \ .
\end{equation}
Defining a effective potential as
\begin{equation}
\overline{U}(f,g_1,..., g_N) =U(f, g_1,..., g_N)-\dfrac{1}{2}\sum_{j=1}^{N} \omega_j^2 g_j^2=\dfrac{1}{2}\Big(f^2+\sum_{j=1}^{N}g_j^2-1\Big)^2+\dfrac{1}{2}\sum_{j=1}^{N}\left(\sigma_j^2-\omega_j^2\right) g_j^2 \label{eq:effective_potential}
\end{equation}
equations (\ref{odes}) can be rewritten as
\begin{equation}\label{eq:FieldEq_EffectivePotential}
    \dfrac{d^2 f}{d x^2}=\dfrac{\partial \overline U}{\partial f} \quad , \quad \dfrac{d^2 g_j}{d x^2}=\dfrac{\partial \overline U}{\partial g_j}
\end{equation}
that can be reinterpreted as a system of Newton equations for a unit mass moving in ${\mathbb R}^{2N+1}$ under a potential $-\overline U$ and with respect to a mechanical time $x$. The corresponding mechanical action functional is thus:
\begin{equation} \label{eq:themechanical_action}
  \overline{E}[f, g_1,..., g_N]=\int_{-\infty}^\infty d x\left( \dfrac{1}{2}\left(\dfrac{d f}{d x}\right)^2+\dfrac{1}{2}\sum_{j=1}^{N}\left(\dfrac{d g_j}{d x}\right)^2+\overline{U}(f, g_1,..., g_N)\right)
\end{equation}
and is related to the energy and the Noether charges of the field theory as
\begin{equation}  \label{eq:rel_energia_accionMecamica}
\begin{aligned}
 &E[f, g_1,..., g_N]=\overline{E}[f, g_1,..., g_N] + \sum_{j=1}^{N} \omega_{j}Q_{j} \ .
\end{aligned}
\end{equation}
Notice that the effective potential (\ref{eq:effective_potential}) has a similar functional form as the original potential function (\ref{eq:Potential_Model}) by simply changing the original model parameters $\sigma_j$ for the new effective model parameters defined by
\begin{equation} \label{eq:effective_model_parameters}
	\Omega_j^2 = \sigma_j^2-\omega_j^2 \ .
\end{equation}
Therefore, making use of (\ref{eq:FieldEq_EffectivePotential}), (\ref{eq:effective_potential}) and (\ref{eq:effective_model_parameters}), we conclude that the profiles $f$ and $g_j$ must comply with the system of differential equations
\begin{equation} \label{eq:generalode}
	\frac{d^2 f}{d x^2}=2 f\left(f^2+g_j^2-1\right)\quad , \quad \frac{d^2 g_j}{d x^2}=2 g_j\left(f^2+g_j^2-1\right)+\Omega_j^2 g_j \ .
\end{equation}
The requirement that vacuum points have to be minimum energy solutions leads now to the inequality $\Omega_{j}^2 > 0$. Having into account (\ref{eq:vacum_hessian}) we obtain the following conditions for the internal rotational frequencies of the $Q$-balls:
\begin{equation} \label{eq:frec1}
   \omega_j^2 < \sigma_j^2 \ .
\end{equation}
The criteria of stability for $Q$-balls was established in \cite{friedberg1976class}. The classically stable solutions under small perturbations are those which minimize the energy at constant Noether charges. In order for this to happen, it is necessary that the second order variation of the energy functional $(\delta^2 E)_{Q_{i}}$ satisfies the condition,
\begin{equation}\label{eq:stability_condicion}
(\delta^2 E)_{Q_{i}} \geq 0
\end{equation}
where $(\delta^2 E)_{Q}$ is given by the expression:
\begin{equation} \label{eq:deltaE}
   (\delta^2 E)_{Q}= \int dx\left(\delta \boldsymbol{\phi}^{t}\mathcal{H} \delta \boldsymbol{\phi}\right) + \sum_{j=1}^{N} \frac{4 \omega_{j}^3}{Q_{j}} \left(\int dx(g_{j}\delta g_{j})\right)^2
\end{equation}
where $\delta \boldsymbol{\phi} = \left(\delta f, \delta g_1, \dots, \delta g_N \right)$ and the Hessian operator $\mathcal{H}$ is defined as:
\begin{equation} \label{eq:HessianMatrix}
    \mathcal{H}=\begin{pmatrix}
-\frac{d^2}{dx^2} + \frac{\partial^2\overline{U}}{\partial f^2}  &  \frac{\partial^2\overline{U}}{\partial f \partial g_{1}} &  \cdots & \frac{\partial^2\overline{U}}{\partial f \partial g_{N}}\\
  \frac{\partial^2\overline{U}}{\partial g_{1} \partial f} & -\frac{d^2}{dx^2} + \frac{\partial^2\overline{U}}{\partial g_{1}^2}  & \cdots   & \frac{\partial^2\overline{U}}{\partial g_{1} \partial g_{N}} \\ \vdots & \vdots & \ddots & \vdots \\
  \frac{\partial^2\overline{U}}{\partial g_{N} \partial f} &  \frac{\partial^2\overline{U}}{\partial g_{N} \partial g_{1}} & \cdots & -\frac{d^2}{dx^2} + \frac{\partial^2\overline{U}}{\partial g_{N}^2}
\end{pmatrix}
\end{equation}
evaluated on the solution.

The study of the stability is in general, a difficult task, and  there does not exist a general criterion to determine whether the condition (\ref{eq:stability_condicion}) is satisfied. It is clear that if the Hessian operator $\mathcal{H}$ has all its eigenvalues positive then (\ref{eq:stability_condicion}) is always satisfied and the solution is stable. An example of this case can be found in \cite{Alonso2023_2}. Another possibility is the existence of only one negative eigenvalue of the Hessian. The argument for the stability in this situation can be determined with a criterion similar to the case where only one complex field is present, studied by Friedberg \textit{et al.} in \cite{friedberg1976class}. Nevertheless, if the number of negative eigenvalues of the Hessian operator is between two and $N-1$, there exists a possibility that the solution to be stable, but the criterion is not yet known. Finally, if the number of negative eigenvalues associated of the Hessian operator is equal to or greater than $N$, condition (\ref{eq:stability_condicion}) is not satisfied and the solution is unstable. However, the criterion \cite{friedberg1976class} can be extended for $N$ complex fields to analyze the stability of the solutions for certain cases. Examples of classical stability criteria for $Q$-balls based on \cite{friedberg1976class} can be found in \cite{ALONSOIZQUIERDO2024}, where non-planar target spaces are considered.

\section{Families of composite defects of kinks and $Q$-balls}\label{Section:3}

To illustrate the defect solutions encompassed by these theories, our focus shifts to the scenario involving $N=2$ complex fields. With two complex scalar fields under $U(1)$ symmetry group transformations, the system presents two distinct conserved Noether charges. This section delves into the influence of these charges on solution structures and their stability. As we carry out this analysis, the extensive complexity of soliton solutions within this system will be unveiled. This specific scenario encapsulates the essence of the general case, offering insights into the behavior of solutions within the broader context of the deformed $O(2N+1)$ nonlinear Sigma model. In the proposed case, the equations (\ref{eq:generalode}) become
\begin{equation} \label{eq:N2ode}
\begin{aligned}
    \dfrac{d^2 f}{d x^2}&= 2f\left(f^2+g_1^2+g_2^2-1\right) \ , \\
    \dfrac{d^2 g_1}{d x^2} &= 2g_1\left(f^2+g_1^2+g_2^2-1\right)+\left( \sigma_{1}^2-\omega_1^2\right)g_{1} \ , \\
    \dfrac{d^2 g_2}{d x^2} &= 2g_2\left(f^2+g_1^2+g_2^2-1\right)+\left( \sigma_{2}^2-\omega_2^2\right)g_{2} \ , \\
\end{aligned}
\end{equation}
where without loss of generality the convention
\begin{equation}\label{convention}
0<\sigma_1^2< \sigma_2^2
\end{equation}
is assumed.

At this point, it is interesting to notice that, in general, the parameters $\Omega_1^2= \sigma_1^2-\omega_1^2$ and $\Omega_2^2= \sigma_2^2-\omega_2^2$ in (\ref{eq:N2ode}) are different. However, if the parameters $\Omega_i$ are fine-tuned such that the condition $\sigma_1^2-\omega_1^2 = \sigma_2^2-\omega_2^2$ is verified, the equations (\ref{eq:N2ode}) acquire a $O(2)$ symmetry in the $g_1$-$g_2$ internal hyperplane. This means that, if the configuration $(f(x),g_1(x),g_2(x))$ is a solution of (\ref{eq:N2ode}) then $(f(x),g_1(x)\cos \alpha,g_2(x)\sin\alpha)$ is a new solution, where $\alpha\in[0,2\pi)$ is a rotation angle in the $g_1$-$g_2$ plane. Consequently, based on the ansatz (\ref{eq:ansatz}), the $Q$-ball type solutions would read as:
\[
( \, f(x) \, , \, g_1(x) \, \cos \alpha \, e^{i\omega_1 t} \, , \, g_2(x) \, \sin\alpha \, e^{i\omega_2 t} \, )
\]
provided that the two $Q_i$-Noether charges are conserved. The mention to this fine-tuning process, which may seem innocuous at this point, will have great significance in the study of the evolution of unstable solutions and the states to which these solutions decay.

As remarked before, (\ref{eq:N2ode}) can be found as the static Euler-Lagrange equations for the effective mechanical action
\begin{equation} \label{action2}
    \overline{E}\left[f,g_1,g_2\right]=\int dx \left[
    \dfrac{1}{2}\left(\dfrac{df}{dx}\right)^2+
    \dfrac{1}{2}\left(\dfrac{dg_1}{dx}\right)^2+
    \dfrac{1}{2}\left(\dfrac{dg_2}{dx}\right)^2+
    \overline{U}\left(f,g_1,g_2\right)
    \right] \ ,
\end{equation}
where now the effective potential reads
\begin{equation} \label{efectivepotential2}
    \overline{U}\left(f,g_1,g_2\right)=
    \dfrac{1}{2}\left(f^2+g_1^2+g_2^2-1\right)^2
    +\dfrac{1}{2}\left(\sigma_1^2-\omega_1^2\right)g_1^2
    +\dfrac{1}{2}\left(\sigma_2^2-\omega_2^2\right)g_2^2  \ .
\end{equation}
From (\ref{eq:rel_energia_accionMecamica}), the action (\ref{action2}) relates to the energy and the Noether $U(1)$ charges via the following expression
\begin{equation} \label{eq:energy_effectiveAction}
  E\left[f,g_1,g_2\right]= \overline{E}\left[f,g_1,g_2\right]+\omega_{1} Q_{1}+ \omega_{2} Q_{2} \ .
\end{equation}
To analytically identify solutions of equations (\ref{eq:N2ode}), we note their functional similarity to the equations examined in \cite{Alonso2000,Alonso2002}, where analytical kink solutions were discovered in a deformed $O(3)$-Sigma model featuring three real scalar fields. Taking advantage of the results of \cite{Alonso2000}, which hinged on the separability of this set of ordinary differential equations in three-dimensional elliptic coordinates, we aim to analytically obtain explicit solutions representing a composite state of a kink and $Q$-balls in the present context.

We anticipate that there are four types of solutions to be studied in detail in the following subsections, categorized by the number of $Q$-balls attached to the kink-like solution associated with the first field component $f$. Note that the $\phi^4$-model is embedded within the model (\ref{eq:Action_Model}). This embedding can be verified by making all field components null except $f$. This means that a pure kink solution with vanishing Noether charges $Q_i=0$, $i=1,2$ is present in this model. Another class of solutions comprises composite defects consisting of a kink and a single $Q$-ball. Within this category, we encounter two distinct types of solutions depending on the complex field (with modulus $g_1$ or $g_2$) that support the $Q$-ball. Additionally, there exists another class of solutions comprising a kink and two individual $Q$-balls. Within this class, three types of solutions are discernible, two of which depend on the complex field supporting the $Q$-balls, while the third type involves both Noether charges $Q_i$ being nonzero. Finally, there exists a more complex type of composite defect wherein the kink is linked with three simple $Q$-balls. These solutions also possess non-zero charges $Q_i$. Our objective in the following subsections is to analytically identify these solutions and investigate their stability in each case. To achieve this, we will adopt the approach introduced by Friedberg, Lee and Sirlin in \cite{friedberg1976class}. In their analysis only fluctuations on the fields that maintain the constancy of the Noether charges are allowed. It will be proved that some of the composite solutions are unstable. Thus, perturbing these configurations might induce forces between the $Q$-balls, ultimately resulting in the destruction of these solutions. The evolution of these unstable objects must maintain the values of the two Noether charges constant, implying that the final state resulting from their evolution is not immediately evident. To explore the trajectory of these unstable solutions, numerical simulations have been conducted for each case. These simulations aim to elucidate the eventual outcome towards which the unstable solutions evolve.

\subsection{Basic entities of the model: defects attaching a $Q$-ball and a kink} \label{sec:basic}

In this section, we will describe the simplest solutions found in this model, which correspond to the basic defects or entities of the system. By this statement, we mean that the rest of the solutions described in the following subsections can be understood as combinations of a finite number of these basic entities distributed throughout space at different points. There are three such basic defects that can be characterized by explicit analytical expressions, allowing for a more in-depth study of their properties. Specifically, we will find a static topological kink and two solutions where a $Q$-ball rotating around one of the axes of the complex field combines with the topological kink. We describe these solutions and their stability in detail in the following items:

\vspace{0.2cm}

\noindent $\bullet$ {\sc $\mathcal{K}$-defects:} The simplest possibility involves considering the model restricted to the orbit
\begin{equation}
    M_1=\{(f,g_1,g_2)\in\mathbb{R}^3:g_1=g_2=0\} \ ,
\end{equation}
which means that all the complex fields vanish. In this case (\ref{eq:N2ode}) reduces to $\dfrac{df}{dx}=\pm\left(f^2-1\right)$, which leads to the kink solution
\begin{equation} \label{kink}
    \mathcal{K}(x)=\left((-1)^{\alpha}\tanh \overline{x} ,0,0\right) \
\end{equation}
where $\overline{x}=x-x_0$ and $x_0$ is the kink center. The values $\alpha=0,1$ allow us to distinguish between the kink and the anti-kink solutions. Note that the first component in (\ref{kink}) corresponds to the kink in the $\phi^4$ model, as previously anticipated. These solutions carry total energy
\begin{equation} \label{eq:energy_tk1_embedded}
E[\mathcal{K}(x)]=\frac{4}{3}
\end{equation}
and vanishing Noether charges, i.e., $Q_i=0$ for $i=1,2$. The analysis of the linear stability of (\ref{kink}) is developed by exerting a small fluctuation on the kink (for each component) and studying its evolution. This scheme leads us to a spectral problem ${\cal H} \xi_n = \lambda_n^2 \xi_n$ where ${\cal H}$ is the second-order small fluctuation operator, which in this case is given by:
\begin{equation} \label{eq:HessianMatrixK1}
    \mathcal{H}[\mathcal{K}(x)]=\begin{pmatrix}
-\dfrac{d^2}{dx^2}-2 + 6 \ \text{tanh}^2 x & 0 & 0\\
 0 & -\dfrac{d^2}{dx^2} -2 + \sigma_1^2 + 2 \ \text{tanh}^2 x  & 0 \\
 0 & 0 & -\dfrac{d^2}{dx^2} -2 + \sigma_2^2 + 2 \ \text{tanh}^2 x
\end{pmatrix} \ .
\end{equation}
Fortunately, the spectral problem in this case decouples into three independent one-dimensional problems, each featuring a Pöschl-Teller potential well. These problems can be analytically solved. The spectrum associated with the longitudinal fluctuations (first entry of (\ref{eq:HessianMatrixK1})) comprises a zero mode with eigenvalue $\overline{\lambda}_0^2=0$, a shape mode with eigenvalue $\overline{\lambda}_1^2 = 3$, and a continuous spectrum emerging at the threshold value $\overline{\lambda}_c^2= 4$. On the other hand, the orthogonal fluctuations (encoded in the second and third entries in (\ref{eq:HessianMatrixK1})) involve spectra consisting of only one discrete mode with eigenvalue $\sigma_i^2-1$ and a continuous spectrum at the value $\sigma_i^2$ for each component $g_i$ with $i=1,2$.

As a consequence, if the coupling constant $\sigma_1^2 \geq 1$, the kink solution (\ref{kink}) is stable because all the eigenvalues of (\ref{eq:HessianMatrixK1}) are non-negative (note the chosen convention (\ref{convention})). In other case, $\sigma_1^2 < 1$ there will be a negative eigenvalue and the kink solution (\ref{kink}) will be an unstable solution, decaying to another state through orthogonal fluctuations. Indeed, these final configurations correspond to static two-component kinks, which can be obtained as particular cases of the solutions which are described in the next items.

\vspace{0.2cm}

\begin{figure}[htbp]
\centering
\subfigure{\includegraphics[width=55mm]{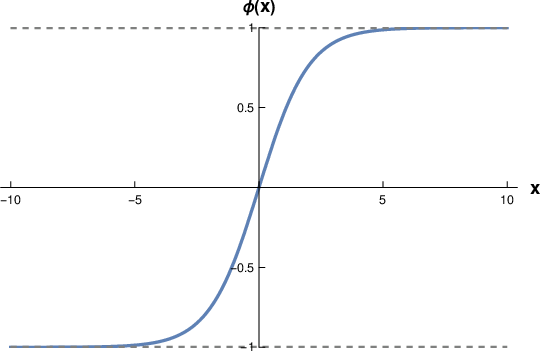}}
\hspace{8mm}
\subfigure{\includegraphics[width=75mm]{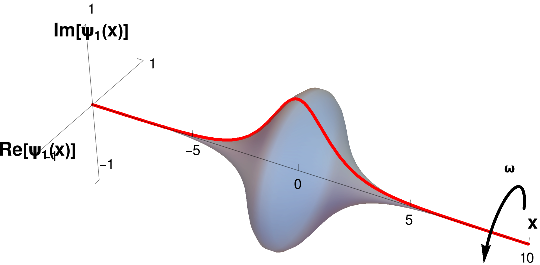}}
\caption{Profiles for the real (left) and the non-vanishing complex (right) components of the $\mathcal{C}_{1}$-defect (\ref{eq:TK2_s1}) with values $\alpha=0$ and $\Omega_1=0.5$.}
 \label{fig:1}
\end{figure}

The next natural step is to search for topological solutions which incorporates not only a non-null $f$-profile but also involves some of the other components $g_i$. From the ansatz (\ref{eq:ansatz}) this defines a $Q$-ball spinning around the axis of the complex field $\psi_i$. Considering that we have incorporated two complex fields in this theory, we expect the presence of two types of the solutions described here. We will now proceed to study these solutions in detail:

\vspace{0.2cm}

\noindent $\bullet$ {\sc $\mathcal{C}_{1}$-defects:} Firstly, the restriction $g_2=0$ will be imposed in equations (\ref{eq:N2ode}), which leads to the reduced equations
\begin{equation} \label{eq:N2odesol1}
\begin{aligned}
    \dfrac{d^2 f}{d x^2}&= 2f\left(f^2+g_1^2-1\right) \ , \\
    \dfrac{d^2 g_1}{d x^2} &= 2g_1\left(f^2+g_1^2-1\right)+\Omega_{1}^2\, g_{1} \ ,
\end{aligned}
\end{equation}
where the notation $\Omega_{1}^2=\sigma_{1}^2-\omega_1^2$ introduced in (\ref{eq:effective_model_parameters}) is explicitly used. Analytical solutions to (\ref{eq:N2odesol1}) have been identified in the range $0<\Omega_{1}^2<1$ \cite{Alonso2023_2}, which leads to a solution for the original equations (\ref{eq:FieldEq_Model}) through the ansatz (\ref{eq:ansatz}). It is described by the expression:
\begin{equation}
 \begin{aligned} \label{eq:TK2_s1}
    \mathcal{C}_{1}(x) = (\phi(x), \psi_1(x),\psi_2(x)) =\left( (-1)^{\alpha} \tanh{(\Omega_{1}\overline{x})}, e^{i \omega_{1} t} \sqrt{1-\Omega_{1}^2} \text{ sech} (\Omega_{1} \overline{x}),0 \right)\ .
 \end{aligned}
\end{equation}
The composite solutions (\ref{eq:TK2_s1}) have been denoted as $\mathcal{C}_{1}(x)$ and they will be briefly referred to as $\mathcal{C}_{1}$-defects. This solution depicts a topological kink, living in the real scalar field $\phi$, coupled with a single $Q$-ball spinning around the axis of the complex field $\psi_1$, see Figure \ref{fig:1}. The parameter $\alpha=0,1$ in  (\ref{eq:TK2_s1}) is used to distinguish between defects and anti-defects, which can be related by a mirror symmetry $x \rightarrow -x$. Note that the presence of the $Q$-ball makes the kink wider than that found in (\ref{kink}). From the construction of (\ref{eq:TK2_s1}), the allowed internal rotational frequencies around the $\psi_1$-axis are restricted to the range
\begin{equation}
    \omega_{1}^2 \in ( \max \{0, \sigma_{1}^2-1 \},\sigma_{1}^2) \ .
\end{equation}
Note that in the case $\sigma_1^2<1$ the $Q$-ball can rotate as slowly as we want, even the spinning can be frozen, $\omega_1^2 = 0$, giving rise to a static two-component kink whose $Q$-charge is equal to zero. Indeed, this would be the solution to which the kink (\ref{kink}) decays in the regime $\sigma_1^2<1$, as previously anticipated.

Substituting the expression (\ref{eq:TK2_s1}) into the Noether charge (\ref{eq:Noether_charges}) and the energy (\ref{eq:Energy}), one obtains respectively these magnitudes for the $\mathcal{C}_{1}$-defects as:
\begin{equation}
 \begin{aligned} \label{eq:Noether_charge_estabilidad_QTK2s1}
   &Q[\mathcal{C}_{1}(x)]= \Big( 2 \omega_{1} \frac{1-\sigma_{1}^2+\omega_{1}^2}{\sqrt{\sigma_{1}^2-\omega_{1}^2}} \, , \, 0 \Big) \ , \\
   &E[\mathcal{C}_{1}(x)]=\frac{2(2\omega_{1}^4-\sigma_{1}^4-\sigma_{1}^2(\omega_{1}^2-3))}{3 \sqrt{\sigma_{1}^2-\omega_{1}^2}} \ .
 \end{aligned}
\end{equation}
The linear stability of the $\mathcal{C}_{1}$-defects (\ref{eq:TK2_s1}) can be determined by analysing the spectrum of the particular Schroedinger-type matrix operator
{\scriptsize
\begin{equation}
\label{eq:HessianMatrixQTK2s1}
    \mathcal{H}[\mathcal{C}_{1}(x)]=	
    \left(
    \begin{array}{ccc}
     -\frac{d^2}{dx^2}+4-2 \left(\Omega_1^2+2\right) \text{sech}^2(\Omega_1 x) & 4 \sqrt{1-\Omega_1^2} \tanh (\Omega_1 x) \text{sech}(\Omega_1 x) & 0 \\
     4 \sqrt{1-\Omega_1^2} \tanh (\Omega_1 x) \text{sech}(\Omega_1 x) & -\frac{d^2}{dx^2}+\left(4-6 \Omega_1^2\right) \text{sech}^2(\Omega_1 x)+\Omega_1^2 & 0 \\
     0 & 0 & -\frac{d^2}{dx^2}+\sigma_2^2-2 \Omega_1^2 \text{sech}^2(\Omega_1 x) \\
    \end{array}
    \right) \ ,
\end{equation}
}
which depends on the parameters $\Omega_1^2$ and $\sigma_2^2$.

\begin{figure}[htbp]
\centering
\includegraphics[width=95mm]{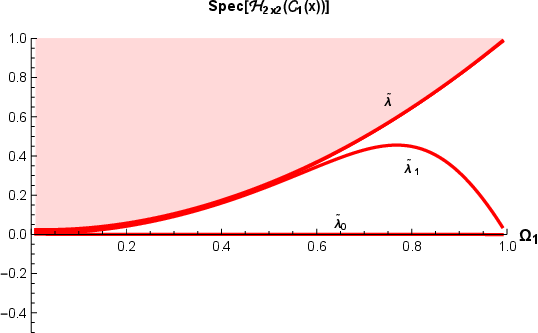}
\caption{Spectrum of the block $\mathcal{H}_{2 \times 2}$ of the second-order small fluctuation operator $\mathcal{H}[\mathcal{C}_{1}(x)]$ (\ref{eq:HessianMatrixQTK2s1}) as a function of the parameter $\Omega_1$.} \label{fig:SpectQTK2_s1}
\end{figure}

Fortunately, the fluctuations in the $g_2$ direction can be decoupled from the others in this spectral problem, leading to a one-dimensional eigenvalue problem with a Pöschl-Teller potential well. The spectrum in this case comprises a discrete eigenvalue $\widetilde{\lambda}_1^2= \sigma_2^2-\Omega_1^2$ and a continuous range $\widetilde{\lambda}^2= \sigma_2^2+k_2^2$, where $k_2\in \mathbb{R}$. Since $\widetilde{\lambda}_1^2=\sigma_{2}^2-\sigma_1^2+\omega_1^2>0$ due to our initial convention (\ref{convention}), we conclude that the $\mathcal{C}_{1}$-defects are stable against fluctuations in the $g_2$-direction. The effect of fluctuations in the $f$ and $g_1$ directions is determined by the first $2 \times 2$ block ${\cal H}_{2\times 2}$ of the Hessian matrix (\ref{eq:HessianMatrixQTK2s1}). Its spectrum must be computed numerically, although it can be shown that $\frac{d\mathcal{C}_{1}(x)}{dx}$ is always a zero mode of (\ref{eq:HessianMatrixQTK2s1}). This zero mode, along with the other eigenvalues associated with this submatrix, are illustrated in Figure \ref{fig:SpectQTK2_s1}. All the eigenvalues that have been found are non-negative regardless of the model parameters. This definitively implies that the solution (\ref{eq:TK2_s1}) is stable. To reinforce this result, numerical simulations have been carried out where the solution has been perturbed in different ways, and the evolution of the solution has been studied. In all cases, it was observed that the solution (\ref{eq:TK2_s1}) does not decay into any other configuration.

\vspace{0.2cm}

\begin{figure}[htbp] \label{Decay01}
\centering
    \subfigure{\includegraphics[width=50mm]{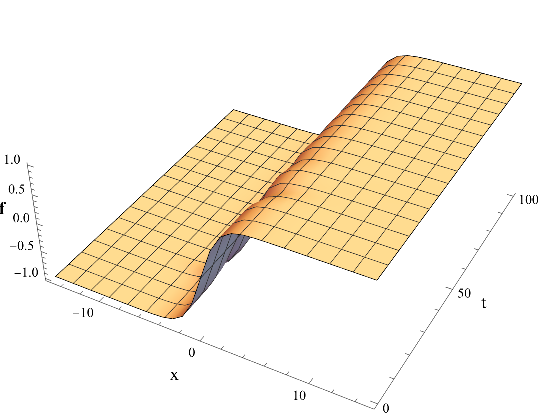}}
    \hspace{4mm}
    \subfigure{\includegraphics[width=50mm]{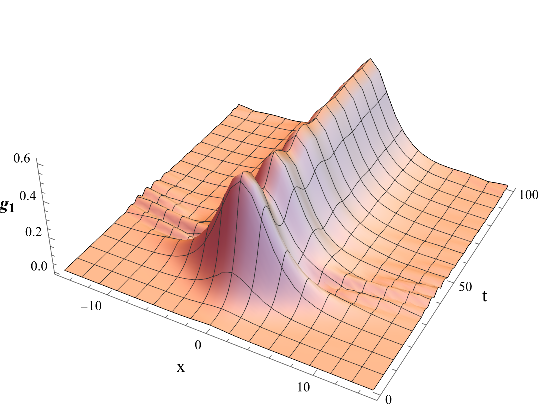}}
    \hspace{4mm}
    \subfigure{\includegraphics[width=50mm]{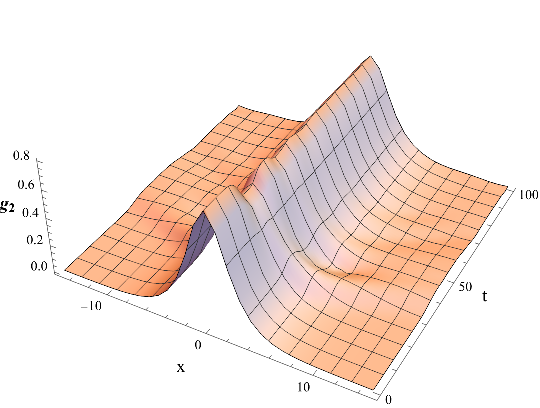}}
\caption{Evolution of the solution $\mathcal{C}_{2}(x)$ perturbed by $\delta g_{1}= {\epsilon}\, \, {\rm sech} ^2{(\Omega_{2}\overline{x})}$ with $\varepsilon=0.01$. The profiles of $f$ (left), $g_{1}$ (middle) and $g_{2}$ (right) are illustrated in this figure for $\sigma_{1}=0.5$, $\sigma_{2}=0.8$, and the initial internal frequencies of the complex fields $\omega_{1}=0$, $\omega_{2}=0.55$. At the end of the simulation, the internal frequencies are $\widetilde{\omega}_{1}=0$ and $\widetilde{\omega}_{2}=0.620852$.}
\end{figure}

\noindent $\bullet$ {\sc $\mathcal{C}_{2}$-defects:} A second class of solutions in this context can be identified if we set the first complex field to zero, $\psi_1=0$, allowing the second one to take non-zero values. Given the symmetry of the model, the expressions characterizing these solutions are easily obtained by interchanging the roles of the complex fields in the expressions provided in (\ref{eq:TK2_s1}) for the $\mathcal{C}_{1}$-defects. In this way, if $0<\Omega_{2}^2<1$, the expressions that determine the now called $\mathcal{C}_{2}$-defects are
\begin{equation}
 \begin{aligned} \label{eq:TK2_s2}
    \mathcal{C}_{2}(x)=\left( (-1)^{\alpha} \tanh{(\Omega_{2}\overline{x})},0, e^{i \omega_{2} t} \sqrt{1-\Omega_{2}^2} \text{ sech} (\Omega_{2} \overline{x}) \right)\ ,
 \end{aligned}
\end{equation}
which now describe a solution composed of a kink in the real field while the Q-ball now rotates around the axis of the second field $\psi_2$. The behavior of these solutions is entirely similar to that of the $\mathcal{C}_{1}$-defects, as shown in Figure \ref{fig:1}, interchanging the roles of the fields $\psi_1$ and $\psi_2$. For the sake of completeness the expression of the Noether charge (\ref{eq:Noether_charges}) and the energy (\ref{eq:Energy}) for the $\mathcal{C}_{2}$-defects are specified as follows
\begin{equation}
 \begin{aligned} \label{eq:energia_carga_tk2_s2}
   &Q[\mathcal{C}_{2}(x)]= \Big( 0 \, , \, 2 \omega_{2} \frac{1-\sigma_{2}^2+\omega_{2}^2}{\sqrt{\sigma_{2}^2-\omega_{2}^2}} \Big) \ , \\
   &E[\mathcal{C}_{2}(x)]=\frac{2\left(2\omega_{2}^4-\sigma_{2}^4-\sigma_{2}^2(\omega_{2}^2-3)\right)}{3 \sqrt{\sigma_{2}^2-\omega_{2}^2}} \ ,
 \end{aligned}
\end{equation}
together with their allowed internal rotation frequencies
\begin{equation} \label{range2}
    \omega_{2}^2 \in (\max\{ 0, \sigma_{2}^2-1\},\sigma_{2}^2) \ .
\end{equation}
Despite these similarities, the results from analysis of the linear stability of these solutions differs. The Hessian operator for this case is given by:
{\scriptsize
\begin{equation} \label{hessian2}
\mathcal{H}[\mathcal{C}_{2}(x)] =	
\left(
\begin{array}{ccc}
 -\frac{d^2}{dx^2}+4-2 \left(\Omega_2^2+2\right) \text{sech}^2(\Omega_2 x) & 0 & 4 \sqrt{1-\Omega_2^2} \tanh (\Omega_2 x) \text{sech}(\Omega_2 x) \\
 0 & -\frac{d^2}{dx^2}+ \sigma_1^2-2 \Omega_2^2 \text{sech}^2(\Omega_2 x) & 0 \\
 4 \sqrt{1-\Omega_2^2} \tanh (\Omega_2 x) \text{sech}(\Omega_2 x) & 0 & -\frac{d^2}{dx^2}+\left(4-6 \Omega_2^2\right) \text{sech}^2(\Omega_2 x)+\Omega_2^2 \\
\end{array}
\right) \ .
\end{equation}
}
As in the previous case, we can decouple the spectral problem associated with the operator (\ref{hessian2}) into different problems: a $2\times 2$ block matrix operator ${\cal H}_{2\times 2}$ can be formed by fluctuations in the $f$ and $g_2$-directions, along with the one-dimensional problem associated with the $g_1$-fluctuations. The first problem leads to a spectrum completely similar to that presented in Figure \ref{fig:SpectQTK2_s1}, involving a zero mode along with positive eigenvalues. This indicates that $\mathcal{C}_{2}$ defects are stable with respect to $f$ and $g_2$ fluctuations. The second spectral problem comprises the discrete eigenvalue $\widehat{\lambda}_1^2= \sigma_{1}^2-\Omega_{2}^2 = \sigma_{1}^2-\sigma_{2}^2+\omega_{2}^2$ together with a continuous spectrum on the threshold value $\sigma_1^2$. Clearly, the sign of the discrete eigenvalue $\widehat{\lambda}_1^2$ depends on the internal rotation frequency $\omega_2$.  It will be positive for $\omega_2^2>\sigma_{2}^2-\sigma_1^2$ and negative for $\omega_2^2<\sigma_{2}^2-\sigma_1^2$. In the first case, it can be concluded that the $\mathcal{C}_{2}$ solution is stable. In the second case, it remains to verify the condition regarding the derivative of the $Q_2$-charge with respect to the frequency $\omega_2$ established in \cite{friedberg1976class}. This condition stipulates that the derivative must be negative to ensure the stability of the solutions (when $\omega_2>0$). Otherwise, the solution is unstable. For $\mathcal{C}_{2}$-defects, it can be checked that
\begin{equation} \label{eq:Nether_s2}
   \frac{d Q_{2}}{d \omega_{2}}>0
\end{equation}
for the range (\ref{range2}) of existence of these solutions. As a final conclusion we can state that the $\mathcal{C}_{2}$-defects are stable if the internal rotation frequency verifies
\[
\omega_2^2 \in \left( \max \{ \sigma_2^2-1, \sigma_{2}^2-\sigma_1^2 \} , \sigma_2^2 \right)
\]
and unstable otherwise. It can be proved that this instability is only possible in the regime where $\sigma_1^2 < 1$, such that the instability regime is given by
\[
\omega_2^2 \in \left( \max \{ 0, \sigma_2^2-1 \}, \sigma_{2}^2-\sigma_1^2 \right) \ .
\]
Numerical simulations have been carried out to validate the previous results. For the cases where the solution is stable, it has been observed that the solution evolves without significant changes. In the range of internal rotation frequencies where the $\mathcal{C}_{2}$-defect is unstable, simulations were performed where the solution (\ref{eq:TK2_s2}) is perturbed by a $g_1$-fluctuation of the type $\delta g_{1}= {\epsilon}\, \, {\rm sech} ^2{(\Omega_{2}\overline{x})}$, corresponding to the eigenfunction with a negative eigenvalue associated with the Hessian (\ref{hessian2}). In these cases, it has been verified that the solutions decay into different configurations. Interestingly, the final configuration to which (\ref{eq:TK2_s2}) evolves in this context is quite surprising. It is found that the $\mathcal{C}_{2}$-solutions change their internal rotation frequency in the second complex field such that the fine-tuning condition $\Omega_1^2 = \Omega_2^2$, that is,
\[
\sigma_1^2 = \sigma_2^2- \widetilde{\omega}_2^2
\]
is satisfied. This implies that the solution accelerates its rotation until reaching the minimum frequency required for stability. As previously mentioned, in this scenario the effective model (\ref{efectivepotential2}) involves the enhancement of symmetry to $O(2)$ in the hyperspace $(g_1,g_2)$. Consequently, the final configuration can be characterized by two non-vanishing profiles for the two complex fields, although the first profile remains static (recall that the first component of the $Q$-charge vanishes) and the second rotates around the $\psi_2$-axis with internal rotation frequency $\widetilde{\omega}_2$. This process of stabilizing the solutions will be referred as \textit{maximal symmetrization mechanism}, where an unstable solution spontaneously looks for the enhancement of the symmetry. Of course, some $Q$-radiation (radiation rotating around the $\psi_2$-axis) is emitted in these processes to keep the value of the second component $Q_2$ constant. It is also possible that some of the internal modes of vibration of these solutions are excited, which means that the profiles exhibit oscillations as they evolve in time, see Figure \ref{Decay01}.

\subsection{Composite defects: combining two basic entities}

\begin{figure}[htbp]
\centering
\begin{tabular}{|c || c c|}
 \hline
   & Real Component ($\phi$) & Complex Component ($\psi_1$) \\ [0.5ex]
 \hline\hline
 \raisebox{2.em}{\rotatebox[origin=l]{90}{$\gamma = 0$}}  &
                 \includegraphics[width=4cm]{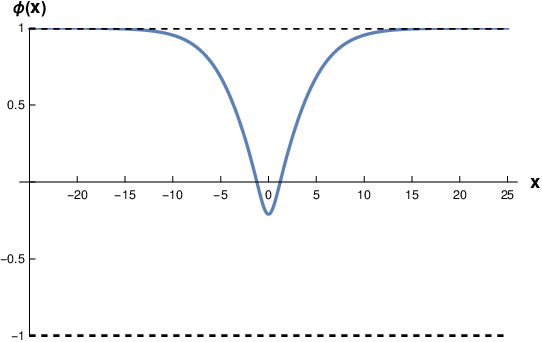} &
                 \includegraphics[width=5cm]{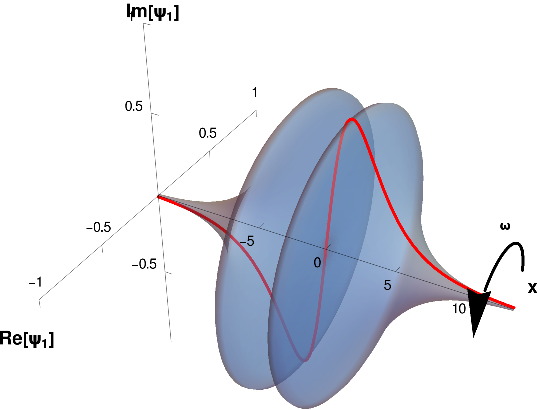}  \\
 \raisebox{2.em}{\rotatebox[origin=l]{90}{$\gamma = 3$}}  &
                 \includegraphics[width=4cm]{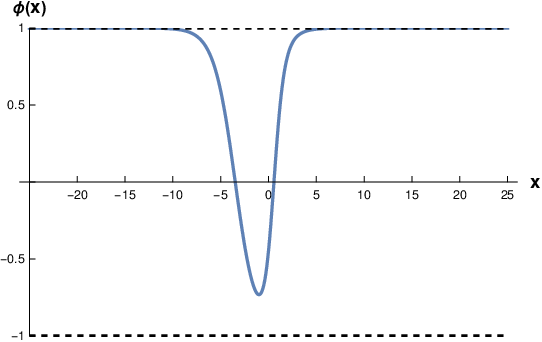} &
                 \includegraphics[width=5cm]{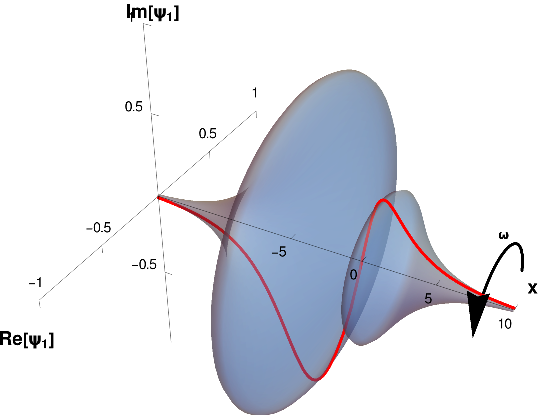}  \\
 \raisebox{2.em}{\rotatebox[origin=l]{90}{$\gamma = 10$}}  &
                 \includegraphics[width=4cm]{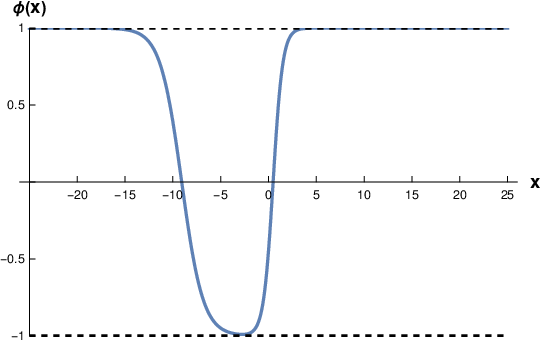} &
                 \includegraphics[width=5cm]{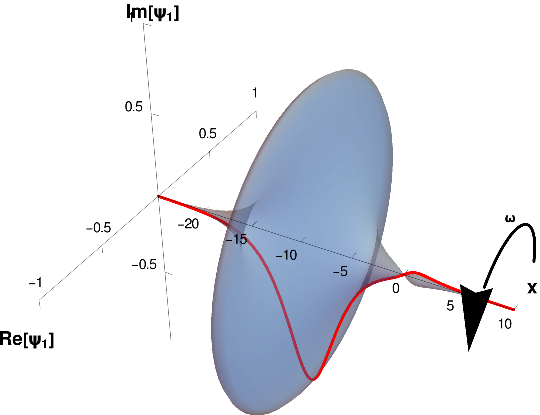}  \\
 \raisebox{2.em}{\rotatebox[origin=l]{90}{$\gamma = 15$}} &
                 \includegraphics[width=4cm]{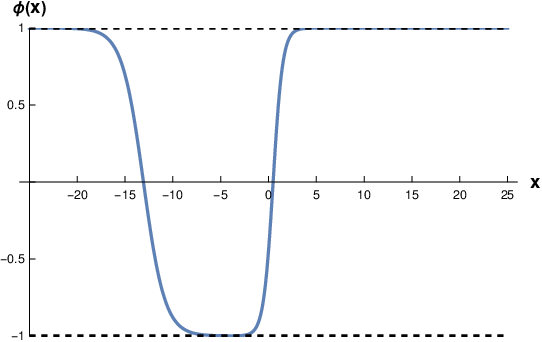} &
                 \includegraphics[width=5cm]{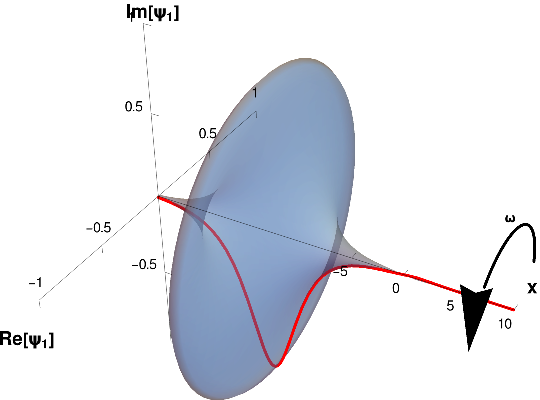} \\
 \hline
\end{tabular}
\caption{Profiles for the real field $\phi$ (left) and the non-vanishing complex component $\psi_1$ (right) of the  $\mathcal{KC}_1(t,x)$ defect (\ref{eq:NTK2}) with values $\alpha=0$, $\sigma_{1}=0.6$, $\omega_{1}=0.4$ and different values of $\gamma$.}
\label{Fig:Fig4}
\end{figure}

In this subsection, we will describe solutions of the model that are constituted by two basic defects (described in the previous section). We will analytically identify all these solutions and study their stability in each case. Specifically, we will find three types of these solutions: two of them combine the topological kink ${\cal K}(x)$ with one of the solutions $\mathcal{C}_i(x)$, while the other combines both types of $\mathcal{C}_i$-defects. They will be described below:

\vspace{0.2cm}

\noindent $\bullet$ {\sc $\mathcal{KC}_1$-defects:} These solutions are characterized by the condition $g_2=0$ and verify the ordinary differential equations (\ref{eq:N2odesol1}). They are determined by the analytical expression
\begin{equation} \label{eq:NTK2a}
    \mathcal{K}\mathcal{C}_{1}(x, t ; \gamma)=\left((-1)^\alpha \frac{\Omega_{-} \cosh \left(\Omega_{+} x_{+}\right)-\Omega_{+} \cosh \left(\Omega_{-} x_{-}\right)}{\Omega_{-} \cosh \left(\Omega_{+} x_{+}\right)+\Omega_{+} \cosh \left(\Omega_{-} x_{-}\right)}, \frac{2 \Omega_{+} \Omega_{-} e^{i \omega_1 t} \sinh \overline{x}}{\Omega_{-} \cosh \left(\Omega_{+} x_{+}\right)+\Omega_{+} \cosh \left(\Omega_{-} x_{-}\right)},0\right) \ ,
\end{equation}
where $\Omega_{\pm}=1 \pm \Omega_{1}$ and $x_{\pm}=\overline{x}-\gamma \Omega_{1}(\Omega_{1} \mp 1)$. As before, the allowed values of the internal rotation frequency are given by
\[
\omega_1^2 \in (\max\{ 0,\sigma_1^2-1 \},\sigma_1^2) \ .
\]
The expression (\ref{eq:NTK2a}) determines a one-parameter family of solutions depending on the parameter $\gamma \in \mathbb{R}$. When the value of $|\gamma|$ is large, the solution describes a composite solution consisting of a topological kink $\mathcal{K}(x)$ and a $\mathcal{C}_2(x)$-defect that are very far apart from each other. Strictly speaking, this configuration is only asymptotically reached for $\gamma \rightarrow \infty$. As the value of $|\gamma|$ begins to decrease, these two preceding basic entities come closer together and progressively distort each other, although they give rise to two energy lumps. However, when $\gamma$ is small, these two basic defects overlap giving rise to only one energy lump. In Figure \ref{Fig:Fig4}, we can see the solutions (\ref{eq:NTK2a}) for four different values of $\gamma$ ($\gamma=0$, $\gamma=3$, $\gamma=10$ and $\gamma=15$), showing the previously mentioned basic defects at distinct separations. Note that although the energy density exhibits only one lump for small values of $\gamma$, the solutions in this regime continues to have two $Q$-ball type profiles in the complex axis $\psi_1$, with one of them shrinking as the value of $|\gamma|$ increases, eventually asymptotically disappearing.

The interpretation above is supported by the sum rule between the energies and the Noether charges of the involved solutions mentioned earlier. We explain this from an analytical point of view. The energy associated with the effective potential $\overline{U}(f,g_1,g_2)$ for this family of solutions verifies the following sum rule (see \cite{Alonso2002}),
\begin{equation} \label{eq:sum_mechanical_action}
    \overline{E}[\mathcal{K}\mathcal{C}_{1} (x,t)]= \overline{E}[\mathcal{K}(x,t)] + \overline{E}[\mathcal{C}_{1} (x,t)] \ .
\end{equation}
From this relation, new sum rules can be obtained for the original field theory. For example, as a consequence of Legendre transformations $\frac{d\overline{E}(\omega_1,\omega_2)}{d\omega_i}=-Q_i$, it can be proved that
\begin{equation} \label{eq:Sum_NoetherCharge}
    \begin{aligned}
    &Q_{1}[\mathcal{K}\mathcal{C}_{1} (x,t)]=Q_{1}[\mathcal{C}_{1}(x,t)] \ , \\
    &Q_{2}[\mathcal{K}\mathcal{C}_{1} (x,t)]=Q_{2}[\mathcal{C}_{1}(x,t)] \ .
    \end{aligned}
\end{equation}
Once that the $Q_i$-charges have been identified, it becomes clear from (\ref{eq:energy_effectiveAction}) that the total energy sum rule
\begin{equation}
      E[\mathcal{K}\mathcal{C}_{1} (x,t)]= E[\mathcal{K}(x,t)] + E[\mathcal{C}_{1} (x,t)]
\end{equation}
holds. This relation between the energies is also valid for other families of solutions (which will be described later), provided that there exists an energy sum rule (such as (\ref{eq:sum_mechanical_action})) associated with the effective potential. In the following subsections we shall only specify the expressions of the $Q$-charges and total energies for the solutions without repeating this argument. In this case we have
\begin{equation}
    \begin{aligned}
      & Q[\mathcal{KC}_1(x,t;\gamma)]=Q[\mathcal{C}_{1}(x,t)]=2 \omega_{1} \frac{1-\sigma_{1}^2+\omega_{1}^2}{\sqrt{{\sigma_1}^2-\omega_{1}^2}} \ , \\
      & E[\mathcal{KC}_1(x,t;\gamma)]=E[\mathcal{K}(x,t)]+E[\mathcal{C}_{1}(x,t)]=\frac{4}{3}+\frac{2(2\omega_{1}^4-\sigma_{1}^4-\sigma_{1}^2(\omega_{1}^2-3))}{3 \sqrt{\sigma_{1}^2-\omega_{1}^2}} \ ,
    \end{aligned}
\end{equation}
which establish that the energy of a $\mathcal{KC}_1$-defect is equal to the sum of the energy of a basic $\mathcal{K}$-defect and a $\mathcal{C}_1$-defect, while its $Q$-charge is determined by the Noether charge of the $\mathcal{C}_1$-defect. Also, it is noteworthy that all members of the family of solutions (\ref{eq:NTK2a}) have the same energy and the same Noether charge, which means that they comprise a family of BPS-like solutions. This allows the adiabatic transition between different members of the family (\ref{eq:NTK2}).

The linear stability of the solutions (\ref{eq:NTK2a}) can be determined by identifying the spectrum of a $3\times 3$ matrix Hessian operator whose non-vanishing entries are given by
\begin{equation}
\begin{aligned} \label{hess3}
\mathcal{H}_{11}[\mathcal{KC}_1(x)]&=-\dfrac{d^2}{dx^2} +\frac{6(\Omega_{-}\cosh{(\Omega_{+}x_{+})}-\Omega_{+}\cosh{(\Omega_{-}x_{-})})^2+8 \Omega_{+}^2\Omega_{-}^2\sinh{(\overline{x})}^2}{(\Omega_{-}\cosh{(\Omega_{+}x_{+})}+\Omega_{+}\cosh{(\Omega_{-}x_{-})})^2} \\
 \mathcal{H}_{12}[\mathcal{KC}_1(x)]&=\mathcal{H}_{21}[\mathcal{KC}_1(x)]= \frac{8 \Omega_{+}\Omega_{-}\left(\Omega_{-}\cosh{(\Omega_{+}x_{+})}-\Omega_{+}\cosh{(\Omega_{-}x_{-})}\right)\sinh{\overline{x}}}{(\Omega_{-}\cosh{(\Omega_{+}x_{+})}+\Omega_{+}\cosh{(\Omega_{-}x_{-})})^2} \\
 \mathcal{H}_{22}[\mathcal{KC}_1(x)]&=-\dfrac{d^2}{dx^2} +\Omega_{2}^2+\frac{2(\Omega_{-}\cosh{(\Omega_{+}x_{+})}-\Omega_{+}\cosh{(\Omega_{-}x_{-})})^2+8 \Omega_{+}^2\Omega_{-}^2\sinh{(\overline{x})}^2}{(\Omega_{-}\cosh{(\Omega_{+}x_{+})}+\Omega_{+}\cosh{(\Omega_{-}x_{-})})^2} \\
\mathcal{H}_{33}[\mathcal{KC}_1(x)]&=-\dfrac{d^2}{dx^2} + \frac{\Omega_{-}^2\left( 2+\Omega_{1}^2\right)\cosh{(\Omega_{+}x_{+})}^2+ 2 \Omega_{+} \Omega_{-} \left(-2 + \Omega_{1}^2\right)\cosh{(\Omega_{+}x_{+})}\cosh{(\Omega_{-}x_{-})}}{(\Omega_{-}\cosh{(\Omega_{+}x_{+})}+\Omega_{+}\cosh{(\Omega_{-}x_{-})})^2} + \\
 &+ \frac{\Omega_{+}^2\left( \left( 2+ \Omega_{1}^2\right)\cosh{(\Omega_{-}x_{-})}^2+24 \Omega_{-}^2\sinh{\overline{x}} \right)}{(\Omega_{-}\cosh{(\Omega_{+}x_{+})}+\Omega_{+}\cosh{(\Omega_{-}x_{-})})^2} \ .
&
\end{aligned}
\end{equation}

\begin{figure}[htbp]
\centering
\includegraphics[width=70mm]{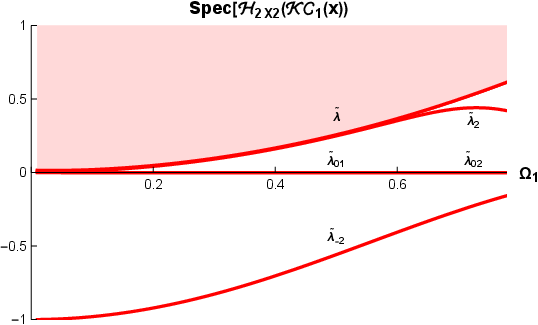}
 \hspace{8mm}
\includegraphics[width=70mm]{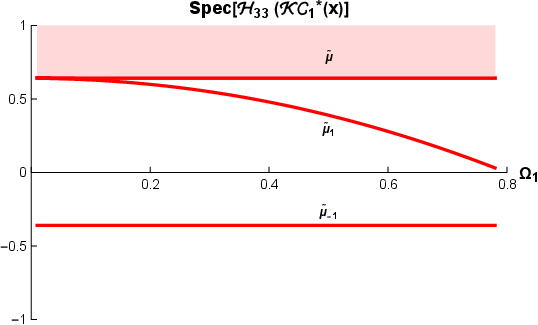}

\includegraphics[width=70mm]{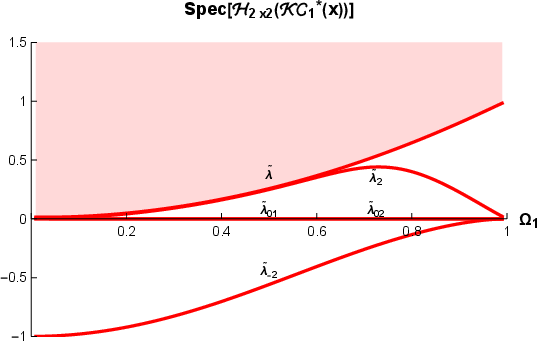}
 \hspace{8mm}
\includegraphics[width=70mm]{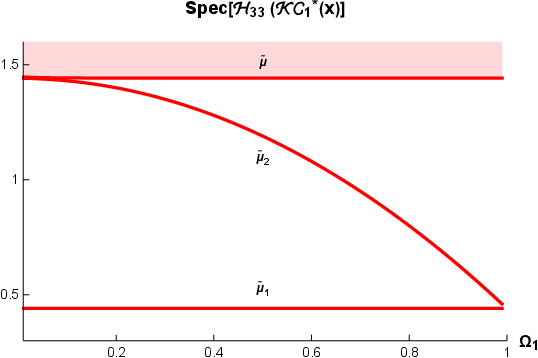}
\caption{Spectrum of the second-order small fluctuation operator
$\mathcal{H}(\mathcal{K}\mathcal{C}_{1})$ as a function of the parameter $\Omega_1$ with $\sigma_{2}=0.8$ (top panel) and $\sigma_{2}=1.2$ (bottom panel) for the family member with $\gamma=0$. The spectrum of the block $\mathcal{H}_{2\times 2}$ is represented in the left and that of the decoupled component $\mathcal{H}_{33}$ in the right. The behaviour is completely similar for the rest of values of $\gamma$.} \label{fig:SpectQTK2_2a}
\end{figure}

In Figure \ref{fig:SpectQTK2_2a} (top panel) the spectrum of the second-order fluctuation operator (\ref{hess3}) is depicted for the value of the coupling constant $\sigma_{2}=0.8$ and the family parameter $\gamma=0$ (representing solutions where the basic defects are overlapped). Qualitatively the same structure is found for other members of the family, included those where the basic defects are far apart. The spectrum is represented as a function of $\Omega_1$. In this case, the presence of two negative eigenvalues is observed, one of them associated with fluctuations in the $f$-$g_1$ plane and the other associated to $g_2$-fluctuations. In Figure \ref{fig:SpectQTK2_2a} (bottom panel) the spectrum is now represented for the value $\sigma_2=1.2$. The main difference is that in this case there is no negative eigenvalue associated with the $g_2$-fluctuations. It can be numerically checked that this is a general pattern. There are two negative eigenvalues for $\sigma_{2}^2<1$, and only one for $\sigma_{2}^2>1$. In addition, the Noether charge verifies that $\frac{\omega_{1}}{Q_{1}}\frac{d Q_{1}}{d \omega_{1}}>0$, which implies that the solution (\ref{eq:NTK2a}) is unstable.

The numerical simulations carried out in this scenario indicate that the decay channels are diverse. For example, for large values of the family parameter $\gamma$ the solution (\ref{eq:NTK2a}) consists of two well separated basic-defects, one of them close to the ${\cal C}_1$-type defect and the other close to the ${\cal K}(x)$-kink solution although with a small perturbation in the $f_1$-component localized at the kink center, which rotates around the $\psi_1$-axis. The main instability channel for $\sigma_1^2<1$ is the decay of the kink to another ${\cal C}_1$-type defect following the mechanism previously commented. However, when $\sigma_1^2\geq 1$, the instability is caused by the frequency change of the small complex component of this part of the solution. On the other hand, when the family parameter $\gamma$ is small and the basic defects are overlapped the instability channel is different. In this case, the solution split into two $C_1$-defects spinning with lower rotation frequency such that the total $Q_1$-charge of the system is conserved while repelling each other (because they are spinning out-of-phase). When an orthogonal fluctuation to the $f-g_1$-plane is applied have been found that it is possible the decay to two $Q$-oscillons.

\begin{figure}[htbp]
\centering
    \subfigure{\includegraphics[width=55mm]{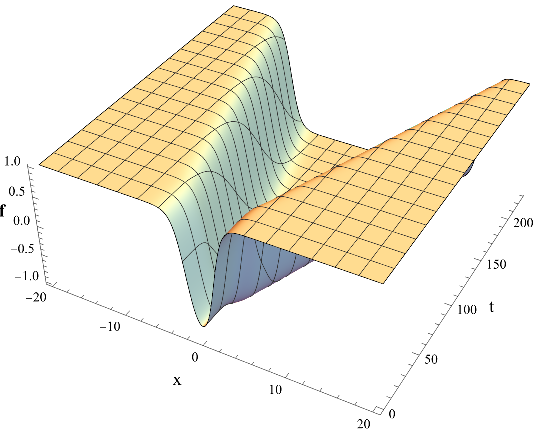}}
    \subfigure{\includegraphics[width=55mm]{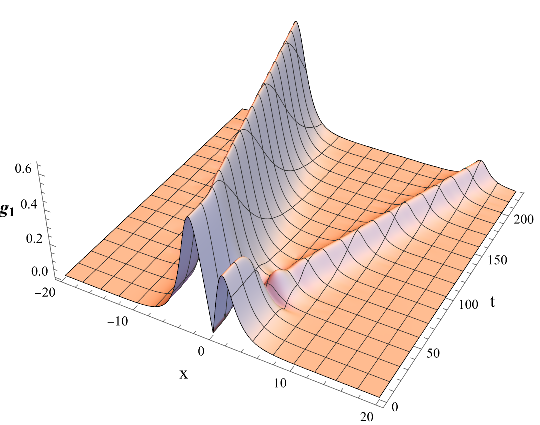}}
    \subfigure{\includegraphics[width=55mm]{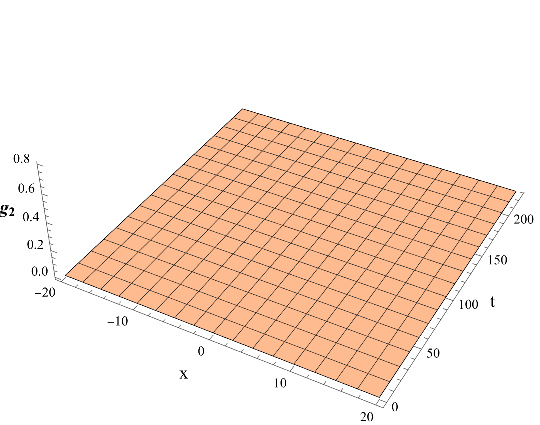}}

\caption{Evolution of the solution $\mathcal{K}\mathcal{C}_{1}(x)$ when perturbed by a fluctuation proportional to the eigenvector associated with the negative eigenvalue of the operator $\mathcal{H}_{2x2}(\mathcal{K}\mathcal{C}_{1})$. The values of the model parameters are $\sigma_{1}=1.1$, $\sigma_{2}=1.2$, the family member is $\gamma=2$ and the initial internal rotation frequency is $\omega_{1}=0.8$. The final internal frequency of defect traveling to the left is $\widetilde \omega=0.797331$ while the frequency of that travelling to the right is $\widetilde \omega=0.475385$.}
\end{figure}

\vspace{0.2cm}

\noindent $\bullet$ {\sc $\mathcal{KC}_2$-defects:} In this case, the solutions are characterized by the condition $g_1 = 0$ and follow the analytical expression
\begin{equation} \label{eq:NTK2}
    \mathcal{KC}_{2}(x, t ; \gamma)=\left((-1)^\alpha \frac{\Omega_{-} \cosh \left(\Omega_{+} x_{+}\right)-\Omega_{+} \cosh \left(\Omega_{-} x_{-}\right)}{\Omega_{-} \cosh \left(\Omega_{+} x_{+}\right)+\Omega_{+} \cosh \left(\Omega_{-} x_{-}\right)},0, \frac{2 \Omega_{+} \Omega_{-} e^{i \omega t} \sinh \overline{x}}{\Omega_{-} \cosh \left(\Omega_{+} x_{+}\right)+\Omega_{+} \cosh \left(\Omega_{-} x_{-}\right)}\right)
\end{equation}
where $\Omega_{\pm}=1 \pm \Omega_{2}$ and $x_{\pm}=\overline{x}-\gamma \Omega_{2}(\Omega_{2} \mp 1)$, and as in the previous subsection, $\alpha=0,1$ distinguish between solutions and anti-solutions. The functional form of the solutions (\ref{eq:NTK2}) is completely similar to that found in (\ref{eq:NTK2a}) and its behaviour can be represented by the Figures contained in the table \ref{Fig:Fig4} considering now that the non-null complex field refers to $\psi_2$ instead of $\psi_1$. Again, the allowed internal rotation frequencies are given by
\[
\omega_{2}^2 \in (\max\{ 0, \sigma_{2}^2-1\},\sigma_{2}^2) \ ,
\]
while the Noether charge $Q$ and the energy $E$ are:
\begin{equation}
    \begin{aligned}
      & Q[\mathcal{KC}_{2}(x,t;\gamma)]=Q[\mathcal{C}_{2}(x,t)]=2 \omega_{2} \frac{1-\sigma_{2}^2+\omega_{2}^2}{\sqrt{{\sigma_2}^2-\omega_{2}^2}} \ , \\
      & E[\mathcal{KC}_{2}(x,t;\gamma)]=E[\mathcal{K}(x,t)]+E[\mathcal{C}_{2}(x,t)]=\frac{4}{3}+\frac{2(2\omega_{2}^4-\sigma_{2}^4-\sigma_{2}^2(\omega_{2}^2-3))}{3 \sqrt{\sigma_{2}^2-\omega_{2}^2}} \ .
    \end{aligned}
\end{equation}
Despite the functional similarity of the solutions, the results from the study of the linear stability is different. The solutions (\ref{eq:NTK2}) can be interpreted as a static kink ${\cal K}(x)$ and a $C_2$-defect located at a distance parameterized by the value of the family parameter $\gamma$, giving rise to two energy lumps. Strictly, the configuration is found for $\gamma \rightarrow \infty$. When $\gamma=0$ these basic defects are completely overlapped and the energy density is localized only around one point. Recall that the ${\cal C}_2$-defect was unstable in the interval $\omega_2^2 \in (\max\{ 0, \sigma_{2}^2-1\}, \sigma_{2}^2-\sigma_{1}^2)$. Also, the static kink is unstable in the range $\sigma_1^2<1$. This makes that the decay channels of this type of solutions are much wider than before. These comments are reflected in the spectrum of the second order small fluctuation operator, whose non-null components are
\begin{equation}
\begin{aligned}
 \mathcal{H}_{11}[\mathcal{K}\mathcal{C}_{2}(x)]&=-\dfrac{d^2}{dx^2} +\frac{6(\Omega_{-}\cosh{(\Omega_{+}x_{+})}-\Omega_{+}\cosh{(\Omega_{-}x_{-})})^2+8 \Omega_{+}^2\Omega_{-}^2\sinh{(\overline{x})}^2}{(\Omega_{-}\cosh{(\Omega_{+}x_{+})}+\Omega_{+}\cosh{(\Omega_{-}x_{-})})^2} \\
 \mathcal{H}_{22}[\mathcal{K}\mathcal{C}_{2}(x)]&=-\dfrac{d^2}{dx^2} +\Omega_{1}^2+\frac{2(\Omega_{-}\cosh{(\Omega_{+}x_{+})}-\Omega_{+}\cosh{(\Omega_{-}x_{-})})^2+8 \Omega_{+}^2\Omega_{-}^2\sinh{(\overline{x})}^2}{(\Omega_{-}\cosh{(\Omega_{+}x_{+})}+\Omega_{+}\cosh{(\Omega_{-}x_{-})})^2} \\
\mathcal{H}_{13}[\mathcal{K}\mathcal{C}_{2}(x)]&= \frac{\Omega_{-}^2\left( 2+\Omega_{2}^2\right)\cosh{(\Omega_{+}x_{+})}^2+ 2 \Omega_{+} \Omega_{-} \left(-2 + \Omega_{2}^2\right)\cosh{(\Omega_{+}x_{+})}\cosh{(\Omega_{-}x_{-})}}{(\Omega_{-}\cosh{(\Omega_{+}x_{+})}+\Omega_{+}\cosh{(\Omega_{-}x_{-})})^2}+\\
&+\frac{ \Omega_{+}^2\left( \left( 2+ \Omega_{2}^2\right)\cosh{(\Omega_{-}x_{-})}^2+24 \Omega_{-}^2\sinh{\overline{x}} \right)}{(\Omega_{-}\cosh{(\Omega_{+}x_{+})}+\Omega_{+}\cosh{(\Omega_{-}x_{-})})^2} \\
\mathcal{H}_{31}[\mathcal{K}\mathcal{C}_{2}(x)]&=\mathcal{H}_{13}[\mathcal{K}\mathcal{C}_{2}(x)] \\
  \mathcal{H}_{33}[\mathcal{K}\mathcal{C}_{2}(x)]&= -\dfrac{d^2}{dx^2} + \frac{8 \Omega_{+}\Omega_{-}\left(\Omega_{-}\cosh{(\Omega_{+}x_{+})}-\Omega_{+}\cosh{(\Omega_{-}x_{-})}\right)\sinh{\overline{x}}}{(\Omega_{-}\cosh{(\Omega_{+}x_{+})}+\Omega_{+}\cosh{(\Omega_{-}x_{-})})^2} \ .
\end{aligned}
\end{equation}
The numerical analysis of the spectrum of the operator ${\cal H}$ indicates that there are three negative eigenvalues for the regime $\Omega_{2}^2>\sigma_{1}^2$ and only two negative eigenvalues for the case $\Omega_{2}^2<\sigma_{1}^2$. As before the criterion of the derivative of the Noether charge leads to the conclusion that this type of solutions are unstable. The numerical simulations carried out in this scenario indicate that the decay channels are similar to those described in the previous solutions. As before for large values of $\gamma$ the constituents of the composite solution (\ref{eq:NTK2}) are well separated and the decay channels are dominated by the behaviour of each of these basic defects. If $\sigma_{1}^2 <1$ the static kink ${\cal K}(x)$ decays to a non-spinning ${\cal C}_1(x)$-defects. At the same time, if $\omega_2^2 \in (\max\{ 0, \sigma_{2}^2-1\}, \max\{ \sigma_{2}^2-1 , \sigma_{2}^2-\sigma_{1}^2\})$ the other constituent, the ${\cal C}_2$-defect, decays following the \textit{maximal symmetrization mechanism}, adopting the internal frequency $\sigma_1^2=\sigma_2^2-\widetilde{\omega}_2^2$, such that the ${\cal C}_2$-defect rotates with a new frequency and such that the other complex field adopts a static non-null profile. For some values of the coupling constants and of the initial internal rotation frequencies these processes may happen simultaneously.

On the other hand, when the family parameter $\gamma$ is small and the basic defects are overlapped we find similar behaviours to the previous case.  In this case, the solution splits into two $C_2$-defects spinning with lower rotation frequency such that the total $Q$-charge of the system is conserved while repelling each other (because they are spinning out-of-phase). In turn, the $C_2$-defects (depending on the regime) can decay into a $C_1$-defect, as previously described. For other phases of the perturbation the simulation ends with a bion, or give place to two $Q$-oscillons.

\vspace{0.2cm}

\begin{figure}[htbp]
\centering
\begin{tabular}{|c || c c c|}
 \hline
   & Real Component ($\phi$) & Complex Component ($\psi_1$) & Complex Component ($\psi_2$)\\ [0.5ex]
 \hline\hline
 \raisebox{2.em}{\rotatebox[origin=l]{90}{$\gamma = 0$}}  &
                 \includegraphics[width=4cm]{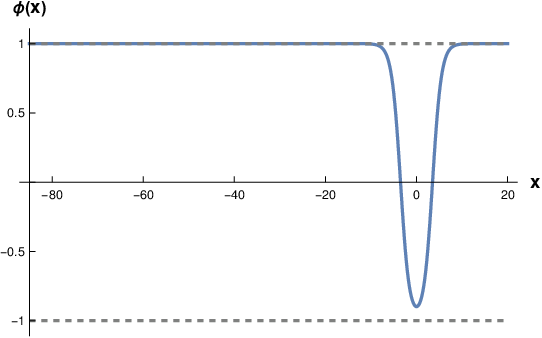} &
                 \includegraphics[width=5cm]{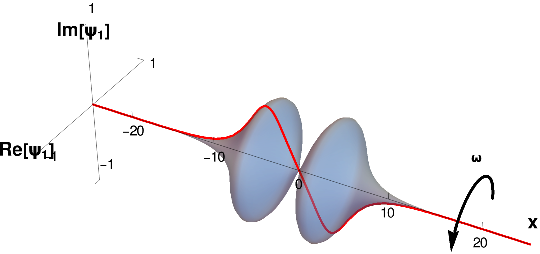} &
                 \includegraphics[width=5cm]{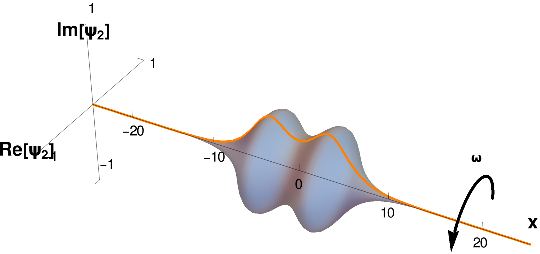}  \\
 \raisebox{2.em}{\rotatebox[origin=l]{90}{$\gamma = 10$}}  &
                 \includegraphics[width=4cm]{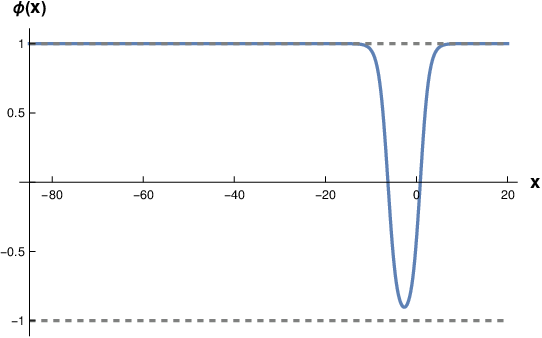} &
                 \includegraphics[width=5cm]{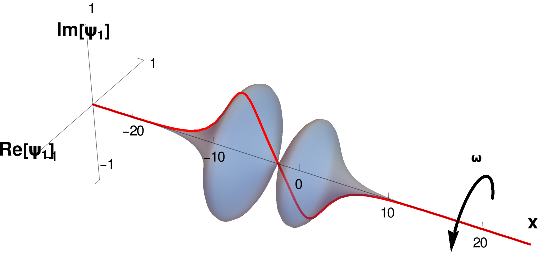} &
                 \includegraphics[width=5cm]{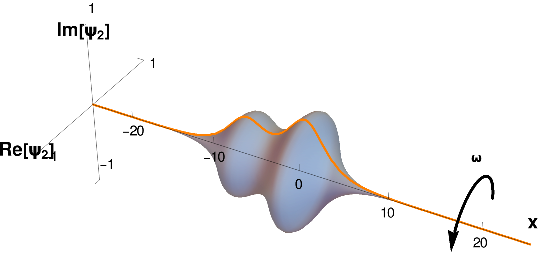}  \\
 \raisebox{2.em}{\rotatebox[origin=l]{90}{$\gamma = 100$}}  &
                 \includegraphics[width=4cm]{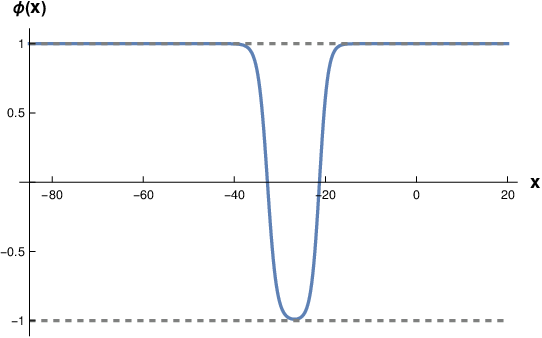} &
                 \includegraphics[width=5cm]{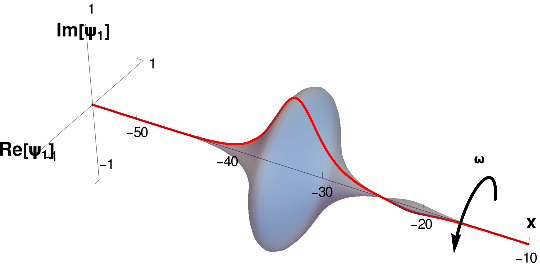} &
                 \includegraphics[width=5cm]{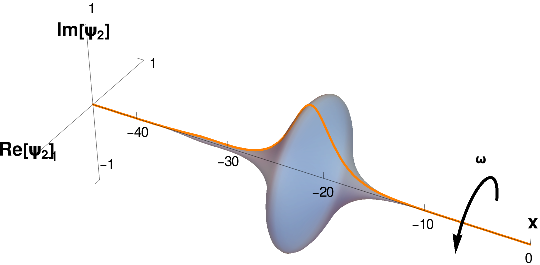}  \\
 \raisebox{2.em}{\rotatebox[origin=l]{90}{$\gamma = 200$}} &
                 \includegraphics[width=4cm]{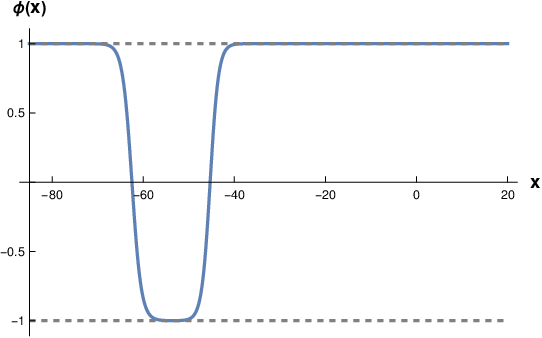} &
                 \includegraphics[width=5cm]{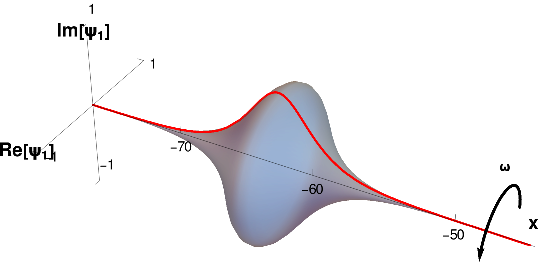} &
                 \includegraphics[width=5cm]{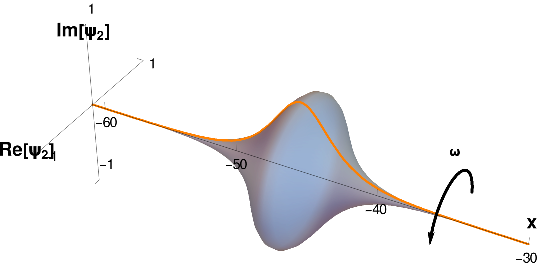} \\
 \hline
\end{tabular}
\caption{Profiles of the solutions $\mathcal{C}_{1}\mathcal{C}_{2} (t,x;\gamma)$
for the particular values of $\sigma_{1}=0.5$, $\omega_{2}=0.1$, $\sigma_{2}=0.6$, $\omega_{2}=0.25$ and different values of $\gamma$ (displayed in each row). The first column illustrates the real component of the solution whereas the second and third columns plot the profile of the complex scalar fields $\psi_1$ and $\psi_2$, respectively.}
\label{Fig:Fig7}
\end{figure}

\noindent $\bullet$ {\sc $\mathcal{C}_1\mathcal{C}_2$-defects:} These solutions are characterized by the analytical expressions
\begin{eqnarray} \label{eq:C1C2}
     \phi(t,x) &= & \frac{\Omega_2 \sinh x_1 \sinh x_2 - \Omega_1 \cosh x_1 \cosh x_2}{\Omega_1 \sinh x_1 \sinh x_2- \Omega_2 \cosh x_1 \cosh x_2 }  \ , \nonumber\\
     \psi_1(t,x)& = & \frac{ \overline{\Omega}_1 \sqrt{\Omega_2^2 - \Omega_1^2} \sinh x_2}{\Omega_1 \sinh x_1 \sinh x_2- \Omega_2 \cosh x_1 \cosh x_2 }e^{i\omega_1 t} \label{eq:QNTK3_QBall_1}\ , \\
     \psi_2(t,x) & = & \frac{-\overline{\Omega}_2 \sqrt{\Omega_2^2 - \Omega_1^2} \cosh x_1}{\Omega_1 \sinh x_1 \sinh x_2- \Omega_2 \cosh x_1 \cosh x_2 } e^{i\omega_2 t} \ , \nonumber
\end{eqnarray}
where $x_1 = \Omega_1 (x+\gamma\,\Omega_2^2)$, $x_2 = \Omega_2 (x+\gamma\,\Omega_1^2)$ and $\overline{\Omega}_i= \sqrt{1-\Omega_{i}^2}$. It is assumed that $\Omega_1^2 <\Omega_2^2$. Of course the other possibility is easy to implement following the previous expression changing the role of the parameters. Note that (\ref{eq:QNTK3_QBall_1}) describe a family of solutions which are characterized by the value of the parameter $\gamma$. When this parameter $|\gamma|$ is large enough the solution (\ref{eq:QNTK3_QBall_1}) corresponds to a ${\cal C}_1(x)$-defect followed by ${\cal C}_2(x)$-defect, which are separated by a certain distance. As before, this interpretation is strict asymptotically when $\gamma \rightarrow \infty$. In other case, the basic defects arise distorted. These ${\cal C}_i$-defects rotate around the $\psi_i$-axis with internal frequency $\omega_i$. When $\gamma=0$ these two defects are completely overlapped, see Figure in table \ref{Fig:Fig7}.

Therefore, the total energy $E$ is given by
\begin{equation}
\begin{aligned}
       E[\mathcal{C}_{1}\mathcal{C}_{2} (x,t)]&= E[\mathcal{C}_{1}(x,t)] + E[\mathcal{C}_{2} (x,t)]  = \\ &=\frac{2(2\omega_{1}^4-\sigma_{1}^4-\sigma_{1}^2(\omega_{1}^2-3))}{3 \sqrt{\sigma_{1}^2-\omega_{1}^2}} + \frac{2(2\omega_{2}^4-\sigma_{2}^4-\sigma_{2}^2(\omega_{2}^2-3))}{3 \sqrt{\sigma_{2}^2-\omega_{2}^2}}  \ ,
\end{aligned}
\end{equation}
while both Noether charges are simultaneously non null for these solutions

\begin{equation}
\begin{aligned}
 &Q_{1}[\mathcal{C}_{1}\mathcal{C}_{2} (x,t)]=Q_{1}[\mathcal{C}_{1}(x,t)]= 2 \omega_{1} \frac{1-\sigma_{1}^2+\omega_{1}^2}{\sqrt{\sigma_{1}^2-\omega_{1}^2}} \ , \\
 &Q_{2}[\mathcal{C}_{1}\mathcal{C}_{2} (x,t)]=Q_{2}[\mathcal{C}_{2}(x,t)]= 2 \omega_{2} \frac{1-\sigma_{2}^2+\omega_{2}^2}{\sqrt{\sigma_{2}^2-\omega_{2}^2}} \ .
\end{aligned}
\end{equation}

As we can see both the energy and the $Q$-charges do not depend on the family parameter $\gamma$, which means that (\ref{eq:C1C2}) corresponds to BPS-like solutions. The Hessian operator ${\cal H}$ for these solutions is determined by the components

\begin{eqnarray*}
\mathcal{H}_{11}[\mathcal{C}_{1}\mathcal{C}_{2} (x)] &=&
- \frac{d^2}{dx^2} +
\frac{2\cosh^2{(x_1)}\left(\left(\Omega_{2}^2-\Omega_{1}^2 \right) \overline{\Omega}_{2}^2+ 3 \Omega_{1}^2 \cosh^2{(x_2)}\right) - 3\Omega_{1} \Omega_{2}\sinh{(2x_{1})}\sinh{(2x_{2})}}{\left(\Omega_{2}\cosh{(x_{1})}\cosh{(x_{2})}-\Omega_{1}\sinh{(x_{1})}\sinh{(x_{2})}\right)^2}\\&+&
\frac{2\sinh^2{(x_2)} \left(\left( \Omega_{2}^2-\Omega_{1}^2 \right) \overline{\Omega}_{1}^2 + 3 \Omega_{2}^2 \sinh^2{(x_1)} \right) }{\left(\Omega_{2}\cosh{(x_{1})}\cosh{(x_{2})}-\Omega_{1}\sinh{(x_{1})}\sinh{(x_{2})}\right)^2} \ , \\ [2mm]
\mathcal{H}_{12}[\mathcal{C}_{1}\mathcal{C}_{2} (x)] &=& \frac{4 \overline{\Omega}_{1}\sqrt{\Omega_{2}^2-\Omega_{1}^2}\sinh{(x_{2})}\left(\Omega_{2}\sinh{(x_{1})}\sinh{(x_{2})}-\Omega_{1}\cosh{(x_{1})}\cosh{(x_{2})}\right)}{\left(\Omega_{2}\cosh{(x_{1})}\cosh{(x_{2})}-\Omega_{1}\sinh{(x_{1})}\sinh{(x_{2})}\right)^2} \ , \\ [2mm]
\mathcal{H}_{13}[\mathcal{C}_{1}\mathcal{C}_{2} (x)] &=& \frac{4 \overline{\Omega}_{2}\sqrt{\Omega_{2}^2-\Omega_{1}^2}\cosh{(x_{1})}\left(\Omega_{1}\cosh{(x_{1})}\cosh{(x_{2})}-\Omega_{2}\sinh{(x_{1})}\sinh{(x_{2})}\right)}{\left(\Omega_{2}\cosh{(x_{1})}\cosh{(x_{2})}-\Omega_{1}\sinh{(x_{1})}\sinh{(x_{2})}\right)^2} \ ,  \\ [2mm]
\mathcal{H}_{22}[\mathcal{C}_{1}\mathcal{C}_{2} (x)] &=& - \frac{d^2}{dx^2} +\frac{\cosh^2{(x_1)}\left( 2 \left( \Omega_{2}^2-\Omega_{1}^2\right) \overline{\Omega}_{2}^2 + \Omega_{1}^2\left(2+ \Omega_{2}^2 \right)\cosh^2{(x_2)}\right)}{\left(\Omega_{2}\cosh{(x_{1})}\cosh{(x_{2})}-\Omega_{1}\sinh{(x_{1})}\sinh{(x_{2})}\right)^2} \\ &+& \frac{\sinh^2{(x_2)}\left( 6 \overline{\Omega}_{1}^2 \left( \Omega_{2}^2-\Omega_{1}^2\right) + \left( \Omega_{1}^4+ 2 \Omega_{2}^2 \right) \sinh^2{(x_1)} \right)}{\left(\Omega_{2}\cosh{(x_{1})}\cosh{(x_{2})}-\Omega_{1}\sinh{(x_{1})}\sinh{(x_{2})}\right)^2} \\ &-& \frac{ \Omega_{1}\Omega_{2} \left( 2+ \Omega_{1}^2 \right) \sinh{(2 x_{1})} \sinh{(2 x_{2})}}{2\left(\Omega_{2}\cosh{(x_{1})}\cosh{(x_{2})}-\Omega_{1}\sinh{(x_{1})}\sinh{(x_{2})}\right)^2} \ , \\  [2mm]
\mathcal{H}_{23}[\mathcal{C}_{1}\mathcal{C}_{2} (x)] &=&\frac{4 \overline{\Omega}_{1}\overline{\Omega}_{2}\left( \Omega_{1}^2-\Omega_{2}^2 \right) \cosh{(x_{1})} \sinh{(x_{2})}}{\left(\Omega_{2}\cosh{(x_{1})}\cosh{(x_{2})}-\Omega_{1}\sinh{(x_{1})}\sinh{(x_{2})}\right)^2} \ , \\
\mathcal{H}_{33}[\mathcal{C}_{1}\mathcal{C}_{2} (x)] &=& - \frac{d^2}{dx^2} +\frac{\cosh^2{(x_1)} \left( 6 \left(\Omega_{2}^2- \Omega_{1}^2 \right)\overline{\Omega}_{2}^2 + \left( \Omega_{2}^4+2 \Omega_{1}^2\right)\cosh^2{x_{2}}\right)}{\left(\Omega_{2}\cosh{(x_{1})}\cosh{(x_{2})}-\Omega_{1}\sinh{(x_{1})}\sinh{(x_{2})}\right)^2} \\ &+& \frac{\sinh^2{(x_2)} \left( 2 \left( \Omega_{2}^2-\Omega_{1}^2\right)\overline{\Omega}_{1}^2 + \left( 2+\Omega_{1}^2\right)\Omega_{2}^2 \sinh^2{(x_1)}\right)}{\left(\Omega_{2}\cosh{(x_{1})}\cosh{(x_{2})}-\Omega_{1}\sinh{(x_{1})}\sinh{(x_{2})}\right)^2} \\ &-& \frac{\Omega_{1} \Omega_{2} \left( 2+ \Omega_{2}^2\right)\sinh{(2 x_1)}\sinh{(2 x_{2})}}{2\left(\Omega_{2}\cosh{(x_{1})}\cosh{(x_{2})}-\Omega_{1}\sinh{(x_{1})}\sinh{(x_{2})}\right)^2} \ , \\ [2mm]
\mathcal{H}_{ij}[\mathcal{C}_{1}\mathcal{C}_{2} (x)]&=&\mathcal{H}_{ji}[\mathcal{C}_{1}\mathcal{C}_{2} (x)] \quad \text{for} \quad i \neq j \ .
\end{eqnarray*}

The spectrum of this Hessian operator has been numerically calculated. There is only one negative eigenvalue. We obtain that $\frac{\omega_{i}}{Q_{i}} \frac{d Q_{i}}{d \omega_{i}}>0$ for $i=1,2$, which means that the solution is unstable. The numerical simulations indicate that there are several decay channels. If the parameter $\gamma$ is large enough, the solution consists of the basic ${\cal C}_i(x)$-defects, so the stability analysis reduces to each of these constituents. In particular, the ${\cal C}_2(x)$-defect can decay in certain regimes following the previously described \textit{maximal symmetrization mechanism}. On the other hand, if $\gamma$ is small, then the solution decays to two defects of ${\cal C}_1{\cal C}_2(x)$ type, which are also defined by \textit{maximal symmetrization mechanism} in such a way that for each of them the condition $\Omega_1=\Omega_2$ is complied with.

\begin{figure}[htbp]
\centering
    \subfigure{\includegraphics[width=55mm]{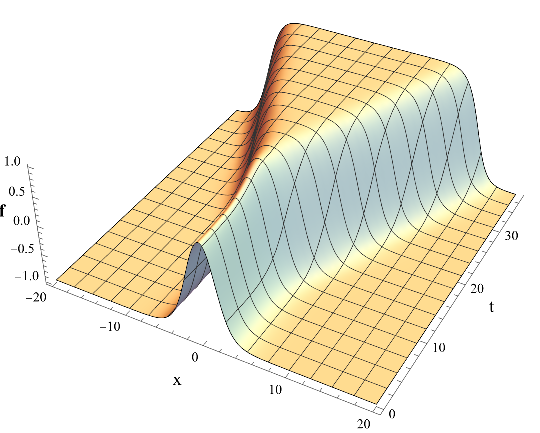}}
    \subfigure{\includegraphics[width=55mm]{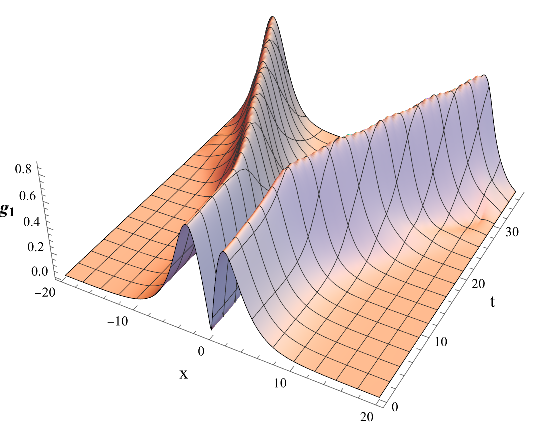}}
    \subfigure{\includegraphics[width=55mm]{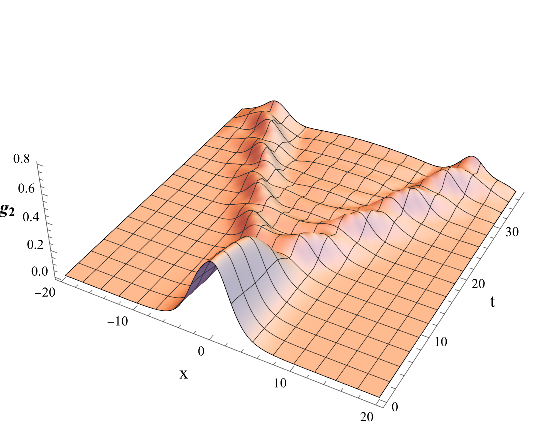}}

\caption{Evolution of the solution $\mathcal{C}_{1}\mathcal{C}_{2}(x)$ with $\gamma=0$ perturbed by a fluctuation proportional to the eigenvector associated with the negative eigenvalue of the Hessian operator $\mathcal{H}(\mathcal{C}_{1}\mathcal{C}_{2})$. The values of the model parameters are $\sigma_{1}=0.5$, $\sigma_{2}=0.8$, the family parameter is $\omega_{1}=0.1$ and the initial internal rotation frequency is $\omega_{2}=0.2$. The final frequencies of the defect travelling to the left are $\widetilde \omega_{1}=0.0995753$ and $\widetilde \omega_{2}=0.631182$ while the frequencies of that travelling to the right are $\widetilde \omega_{1}=0.00293045$ and $\widetilde \omega_{2}=0.620645$. It can be checked that $\widetilde \Omega_{1}^2 \approx \widetilde \Omega_{2}^2$ in both right and left topological defects. }
\end{figure} \label{fig:C1C2_gamma_1_s1_s2_menor_1}

\subsection{Composite defects: combining three basic entities}

\begin{figure}[htbp]
    \centering
    \subfigure{\includegraphics[width=45mm]{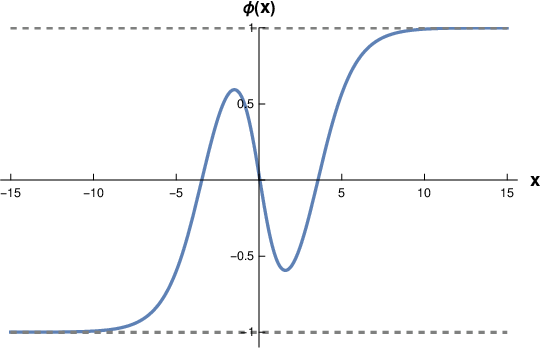}}
    \subfigure{\includegraphics[width=55mm]{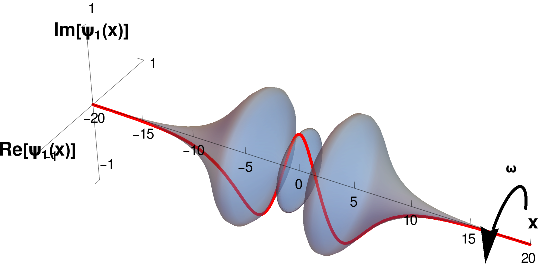}}
    \subfigure{\includegraphics[width=55mm]{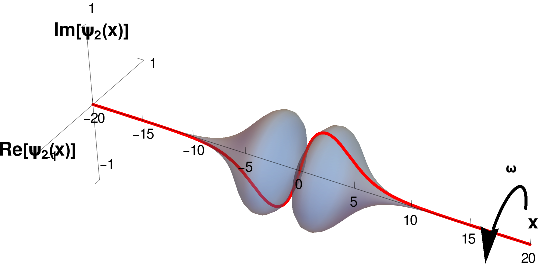}}
    \caption{Graphics of the solution $\mathcal{C}_{1}\mathcal{K}\mathcal{C}_{2}(t,x)$ composed by a topological Kink in the real component and two $Q$-ball for the particular values $\Omega_{1}=0.4$, $\Omega_{2}=0.6$ with $\gamma_{2}=\gamma_{3}=0$.}
     \label{fig:QTK3}
\end{figure}

In addition to the previous composite defects, there exist other families of solutions involving the combination of three of the basic defects introduced in Section \ref{sec:basic}. Unfortunately, it is not possible to find explicit analytical expressions for the solutions in this case, although we can obtain implicit ones. This can be got by using a system of elliptic coordinates, defined as follows:
\begin{eqnarray*}
f^2&=&\frac{\left(1-\lambda_1\right)\left(1-\lambda_2\right)\left(1-\lambda_3\right)}{\Omega_2^2 \Omega_3^2}, \\
g_1^2&=&\frac{\left(\overline{\Omega}_2^2-\lambda_1\right)\left(\overline{\Omega}_2^2-\lambda_2\right)\left(\overline{\Omega}_2^2-\lambda_3\right)}{-\Omega_2^2\left(\Omega_3^2-\Omega_2^2\right)}, \\
g_2^2&=&\frac{\left(\overline{\Omega}_3^2-\lambda_1\right)\left(\overline{\Omega}_3^2-\lambda_2\right)\left(\overline{\Omega}_3^2-\lambda_3\right)}{\Omega_3^2\left(\Omega_3^2-\Omega_2^2\right)},
\end{eqnarray*}
where $\overline{\Omega}_{i}^2= 1-\Omega_{i}^2$. In these variables the system of ordinary differential equations (\ref{eq:N2ode}) can be solved, leading to the implicit relations
\begin{equation}\label{tk31}
 \begin{aligned}
 C_1(\tau)=&\left|\frac{\sqrt{1-\lambda_1}-\Omega_2}{\sqrt{1-\lambda_1}+\Omega_2}\right|^{\Omega_3 \overline{\Omega}_2^2 \operatorname{sign}\left(\pi_1\right)}\left|\frac{\sqrt{1-\lambda_1}+\Omega_3}{\sqrt{1-\lambda_1}-\Omega_3}\right|^{\Omega_2 \overline{\Omega}_3^2 \operatorname{sign}\left(\pi_1\right)} \\
& \times\left|\frac{\sqrt{1-\lambda_2}+\Omega_2}{\sqrt{1-\lambda_2}-\Omega_2}\right|^{\Omega_3 \overline \Omega_2^2 \operatorname{sign}\left(\pi_2\right)}\left|\frac{\sqrt{1-\lambda_2}-\Omega_3}{\sqrt{1-\lambda_2}+\Omega_3}\right|^{\Omega_2 \overline \Omega_3^2 \operatorname{sign}\left(\pi_2\right)} \\
& \times\left|\frac{\sqrt{1-\lambda_3}-\Omega_2}{\sqrt{1-\lambda_3}+\Omega_2}\right|^{\Omega_3 \overline{\Omega}_2^2 \operatorname{sign}\left(\pi_3\right)}\left|\frac{\sqrt{1-\lambda_3}+\Omega_3}{\sqrt{1-\lambda_3}-\Omega_3}\right|^{\Omega_2 \overline{\Omega}_3^2 \operatorname{sign}\left(\pi_3\right)} \ , \\
&
 \end{aligned}
\end{equation}
where $C_1(x)=e^{2(\gamma_{1}+x)(\Omega_{3}^2-\Omega_{2}^2)\Omega_{2}^2\Omega_{3}^2}$,
 \begin{equation}\label{tk32}
   \begin{aligned}
 C_2=&\left|\frac{\sqrt{1-\lambda_1}+\Omega_2}{\sqrt{1-\lambda_1}-\Omega_2}\right|^{\Omega_3 \operatorname{sign}\left(\pi_1\right)}\left|\frac{\sqrt{1-\lambda_1}-\Omega_3}{\sqrt{1-\lambda_1}+\Omega_3}\right|^{\Omega_2 \operatorname{sign}\left(\pi_1\right)} \\
& \times\left|\frac{\sqrt{1-\lambda_2}+\Omega_2}{\sqrt{1-\lambda_2}-\Omega_2}\right|^{\Omega_3 \operatorname{sign}\left(\pi_2\right)}\left|\frac{\sqrt{1-\lambda_2}-\Omega_3}{\sqrt{1-\lambda_2}+\Omega_3}\right|^{\Omega_2 \operatorname{sign}\left(\pi_2\right)} \\
& \times\left|\frac{\sqrt{1-\lambda_3}+\Omega_2}{\sqrt{1-\lambda_3}-\Omega_2}\right|^{\Omega_3 \operatorname{sign}\left(\pi_3\right)}\left|\frac{\sqrt{1-\lambda_3}-\Omega_3}{\sqrt{1-\lambda_3}+\Omega_3}\right|^{\Omega_2 \operatorname{sign}\left(\pi_3\right)} \ , \\
&
 \end{aligned}
\end{equation}
where $C_{2}=e^{2 \gamma_{2}\Omega_{2}\Omega_{3}(\Omega_{2}^2-\Omega_{3}^2)}$, and
\begin{equation}\label{tk33}
\begin{aligned}
 C_3=&\left|\frac{\sqrt{1-\lambda_1}-1}{\sqrt{1-\lambda_1}+1}\right|^{\Omega_2 \Omega_3\left(\Omega_2^2-\Omega_3^2\right) \operatorname{sign}\left(\pi_1\right)}\left|\frac{\sqrt{1-\lambda_1}+\Omega_2}{\sqrt{1-\lambda_1}-\Omega_2}\right|^{\Omega_3 \overline{\Omega}_3^2 \operatorname{sign}\left(\pi_1\right)} \\
& \times\left|\frac{\sqrt{1-\lambda_1}-\Omega_3}{\sqrt{1-\lambda_1}+\Omega_3}\right|^{\Omega_2 \overline{\Omega}_2^2 \operatorname{sign}\left(\pi_1\right)}\left|\frac{\sqrt{1-\lambda_2}-1}{\sqrt{1-\lambda_2}+1}\right|^{\Omega_2 \Omega_3\left(\Omega_2^2-\Omega_3^2\right) \operatorname{sign}\left(\pi_2\right)} \\
& \times\left|\frac{\sqrt{1-\lambda_2}+\Omega_2}{\sqrt{1-\lambda_2}-\Omega_2}\right|^{\Omega_3 \overline{\Omega}_3^2 \operatorname{sign}\left(\pi_2\right)}\left|\frac{\sqrt{1-\lambda_2}-\Omega_3}{\sqrt{1-\lambda_2}+\Omega_3}\right|^{\Omega_2 \overline{\Omega}_2^2 \operatorname{sign}\left(\pi_2\right)} \\
& \times\left|\frac{\sqrt{1-\lambda_3}-1}{\sqrt{1-\lambda_3}+1}\right|^{\Omega_2 \Omega_3\left(\Omega_2^2-\Omega_3^2\right) \operatorname{sign}\left(\pi_3\right)}\left|\frac{\sqrt{1-\lambda_3}+\Omega_2}{\sqrt{1-\lambda_3}-\Omega_2}\right|^{\Omega_3 \overline{\Omega}_3^2 \operatorname{sign}\left(\pi_3\right)} \\
& \times\left|\frac{\sqrt{1-\lambda_3}-\Omega_3}{\sqrt{1-\lambda_3}+\Omega_3}\right|^{\Omega_2 \overline{\Omega}_2^2 \operatorname{sign}\left(\pi_3\right)}  \ , \\
&
\end{aligned}
\end{equation}
where $C_{3}=e^{2 \gamma_{3}\Omega_{2}\Omega_{3} \overline{\Omega}_{2}^2\overline{\Omega}_{3}^2(\Omega_{2}^2-\Omega_{3}^2)}$ and $\pi_i = \frac{d\lambda_i}{dx}$. As the previous cases, the frequencies are restricted by $\omega_{1}^2 \in ( \max \{0,\sigma_{1}^2-1\},\sigma_{1}^2)$ and $\omega_{2}^2 \in ( \max \{ 0, \sigma_{2}^2-1 \},\sigma_{2}^2)$. The relations (\ref{tk31}), (\ref{tk32}) and (\ref{tk33}) define the profiles $f$ and $g_i$ as a function of the spatial coordinate $x$ depending on the three parameters $\gamma_i$, $i=1,2,3$. The parameter $\gamma_1$ fixes the defect center while the others determine the member of this family of solutions. These solutions can be understood as a combination of the three basic defects described in section \ref{sec:basic}, which appear well-separated for large values of the parameters $\gamma_i$, $i=2,3$. We shall call these solutions as ${\cal C}_1{\cal K}{\cal C}_2$-defects, emphasizing the fact that these solutions consist of the three basic entities. In Figure  \ref{fig:QTK3}, the real and complex scalar field profiles for the member $\gamma_2=\gamma_3=0$ have been plotted for the values $\Omega_1=0.4$ and $\Omega_2=0.6$, where we can see the internal structure of this type of solutions.

Despite the fact that we have only the implicit expressions for these solutions their energy can be obtained explicitly,
\begin{equation} \label{eq:energy_tk3_generic}
\begin{aligned}
E[\mathcal{C}_{1}\mathcal{K}\mathcal{C}_{2}(x,t)]  &= E[\mathcal{C}_{1}(x,t] +   E[\mathcal{K}(x,t)]+   E[\mathcal{C}_{2}(x,t)] = \\ & = \frac{4}{3} + \frac{2(2\omega_{1}^4-\sigma_{1}^4-\sigma_{1}^2(\omega_{1}^2-3))}{3 \sqrt{\sigma_{1}^2-\omega_{1}^2}} + \frac{2(2\omega_{2}^4-\sigma_{2}^4-\sigma_{2}^2(\omega_{2}^2-3))}{3 \sqrt{\sigma_{2}^2-\omega_{2}^2}} \ ,
\end{aligned}
\end{equation}
which supports the previous interpretation of this family of solutions as a combination of the three basic defects. In addition to this, the Noether charges verify
\begin{equation}
  Q_{1}[\mathcal{C}_{1}\mathcal{K}\mathcal{C}_{2}(x,t)]=Q_{1}[\mathcal{C}_{1}(x,t)]\ , \hspace{2cm}
  Q_{2}[\mathcal{C}_{1}\mathcal{K}\mathcal{C}_{2}(x,t)]=Q_{2}[\mathcal{C}_{2}(x,t)] \ .
\end{equation}
Because these solutions have three constituents, the number of decay channels is widely diverse. Each of these basic lumps can decay by the channels described in the previous subsections, destabilising the composite solutions. In addition to this, for small $\gamma_i$ ($i=2,3$), when all the basic defects are overlapped, the solution decays to a $\mathcal{C}_{1}\mathcal{C}_{2}$-defect by following the \textit{maximal symmetrization mechanism}, where the final configuration is determined by the condition $\Omega_{1}^2 \approx \Omega_{2}^2$. This process involves the annihilation of a defect-anti-defect pair giving rise to radiation pulses moving apart from the origin. Other decay channels corresponds to the formation of bions between the two defects formed after the splitting of the original composite defect.

\begin{figure}[htbp]
    \centering
    \subfigure{\includegraphics[width=55mm]{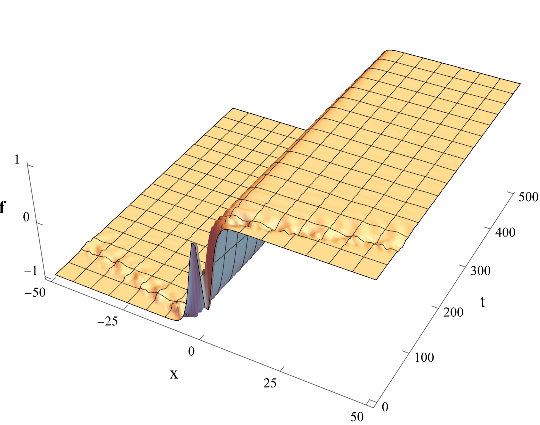}}
    \subfigure{\includegraphics[width=55mm]{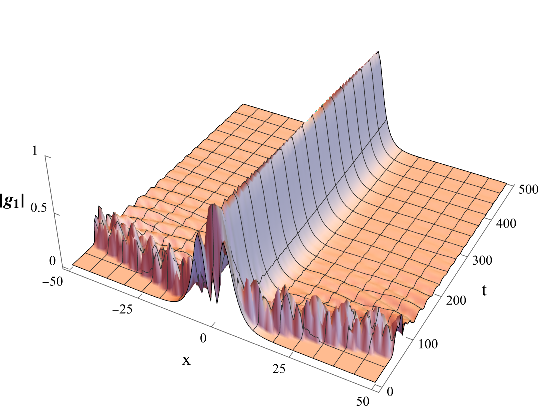}}
    \subfigure{\includegraphics[width=55mm]{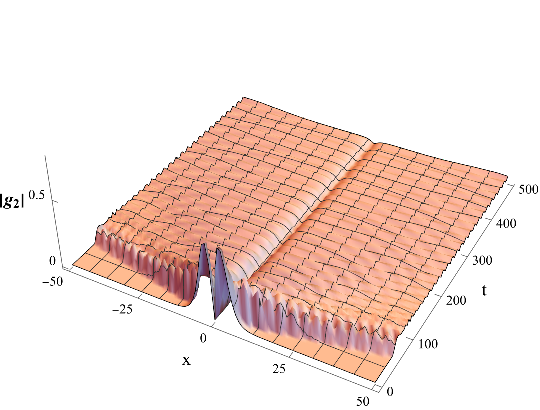}}
    \caption{Evolution of the solution $\mathcal{C}_{1}\mathcal{K}\mathcal{C}_{2}(t,x)$  for the particular initial values $\omega_{1}=0.3$, $\omega_{2}=0.52915$, $\Omega_{1}=0.4$, $\Omega_{2}=0.6$, $\sigma_{1}=0.5$, $\sigma_{2}=0.8$, $\gamma_{2}=0$ and $\gamma_{3}=0$. At the end of the simulation $\Omega_{1}\approx \Omega_{2}$ with $\widetilde \omega_{1}=0.275711$, $\widetilde \omega_{2}=-0.683719$, $\widetilde \Omega_{1}=0.417113$ and $\widetilde \Omega_{2}=0.415366$}
     \label{fig:8}
\end{figure}

\section{Conclusions and further comments}\label{Section:4}

In this article, we have studied the structure of defect type solutions that can be found in a $(1+1)$-dimensional deformed $O(2N+1)$ Sigma model with one real and $N$ complex scalar fields. This model involves a $U(1)$ internal symmetry for each of the complex fields, giving rise to $N$ (Noether) $Q$-charges. We have analyzed the particular $O(5)$-Sigma model where two $Q$-charges are present. The reason is that the patterns of solutions in these models are repeated as the number of complex fields increases. For this particular case, we have been able to find the analytical expression of all the defect type solutions, obtaining also the analytical expression of their energies and $Q$-charges. The structure of the set of all this solutions is also very interesting. There are three single solutions which we have been referred to as \textit{basic defects} because they are the simplest ones that can be found. One of this solution corresponds to a static kink ${\cal K}(x)$ similar to that found in the $\phi^4$-model and which carries out a topological charge. The other two ${\cal C}_1(x)$ and ${\cal C}_2(x)$ correspond to solutions which combine a kink-type solution in the real field and a $Q$-ball in one of the two complex fields, given place to a non-null $Q$-charge in that component. These solutions carry, therefore, topological and Noether charges. In addition to these, there are families of composite solutions which can be understood as a combination of the previously mentioned basic defects. There are families ${\cal KC}_1(x)$, ${\cal KC}_2(x)$ and ${\cal C}_1{\cal C}_2(x)$ combining two different basic defects but there are also families ${\cal C}_1{\cal KC}_2(x)$ combining the three at the same time.

The stability of these solutions have been analyzed following the criterion established by Friedberg, Lee and Sirlin in \cite{friedberg1976class}. This analysis has revealed that the kink-type solution ${\cal K}(x)$ is stable for the parameter range $\sigma_1^2>1$, the ${\cal C}_1$-defect is always stable while the ${\cal C}_2$-defect is stable when the rotation frequency is greater than a critical value $\omega_c$ in the range of possible internal rotation frequencies but unstable in other case. This stability criterion is then tested in numerical simulations where we perturbed the obtained analytical solutions. One novelty is the identification of the decay channels of these solutions. It turns out that the evolution of the ${\cal C}_2(x)$-defect is driven by the \textit{maximal symmetrization mechanism}, where the deformed $O(5)$-symmetry is changed to be exact in the effective model, that is, the final frequencies of the complex field profiles are characterized by the condition $\Omega_1^2 = \Omega_2^2$. In this situation, the ${\cal C}_2(x)$-defect reaches the lowest frequency needed to be stable.

In the case of more complicated, compound solutions, we also identified stability regimes. Interestingly, for unstable solutions, the actual decay channel depends not only on the model parameters but also on the value of the moduli-like parameter $\gamma$. Again, the \textit{maximal symmetrization mechanism} seems to be the main driven force governing the properties of the final state.

This work opens future prospects to be investigated. First of all one should explain the \textit{maximal symmetrization mechanism}. Although observed in various decays, the origin of this synchronization leading to the enhancement of symmetry is rather unclear. One may begin with verification of its existence in other $Q$-ball supporting models. This requires at least two complex scalars. It would also be interesting to study this mechanism in models with bigger number of complex fields.

Another interesting direction is detailed understanding of decay channels of unstable $Q$-balls.

 \section*{Acknowledgements}
This research was funded by the Spanish Ministerio de Ciencia e Innovaci\'on (MCIN) with funding from the European Union NextGenerationEU (PRTRC17.I1) and the Consejer\'{\i}a de Educaci\'on, Junta de Castilla y Le\'on, through QCAYLE project, as well as grant PID2020-113406GB-I00 MTM  funded by MCIN/AEI/10.13039/501100011033.


\begin{thebibliography}{99}
\bibitem{manton2004topological}
N.~Manton and P.~Sutcliffe.
\newblock {\em Topological solitons}.
\newblock Cambridge University Press, 2004.

\bibitem{shnir_2018}
Y.~M. Shnir.
\newblock {\em Topological and Non-Topological Solitons in Scalar Field Theories}.
\newblock Cambridge Monographs on Mathematical Physics. Cambridge University Press, 2018.

\bibitem{friedberg1976class}
R.~Friedberg, T.D. Lee, and A.~Sirlin.
\newblock Class of scalar-field soliton solutions in three space dimensions.
\newblock {\em Physical Review D}, 13(10):2739, 1976.

\bibitem{coleman1985q}
S.~Coleman.
\newblock ${Q}$-balls.
\newblock {\em Nuclear Physics B}, 262(2):263--283, 1985.

\bibitem{coleman1986errata}
S.~Coleman.
\newblock Errata: ${Q}$-balls.
\newblock {\em Nucl. Phys. B}, 269:744, 1986.

\bibitem{LEE1992}
T.D. Lee and Y.~Pang.
\newblock Nontopological solitons.
\newblock {\em Physics Reports}, 221(5):251--350, 1992.

\bibitem{Dine2003}
M.~Dine and A.~Kusenko.
\newblock Origin of the matter-antimatter asymmetry.
\newblock {\em Rev. Mod. Phys.}, 76:1--30, Dec 2003.

\bibitem{tsumagari2008some}
M.I. Tsumagari, E.J. Copeland, and P.M. Saffin.
\newblock Some stationary properties of a ${Q}$-ball in arbitrary space dimensions.
\newblock {\em Physical Review D}, 78(6):065021, 2008.

\bibitem{derrick1964comments}
G.H. Derrick.
\newblock Comments on nonlinear wave equations as models for elementary particles.
\newblock {\em Journal of Mathematical Physics}, 5(9):1252--1254, 1964.

\bibitem{Optics}
M.~J. Ablowitz and J.~T. Cole.
\newblock Nonlinear optical waveguide lattices: Asymptotic analysis, solitons, and topological insulators.
\newblock {\em Physica D: Nonlinear Phenomena}, 440:133440, 2022.

\bibitem{Mollenauer2006}
L.F. Mollenauer and J.P. Gordon.
\newblock {\em Solitons in optical fibers - Fundamentals and applications}.
\newblock Academic Press, 2006.

\bibitem{Schneider2004}
T.~Schneider.
\newblock {\em Nonlinear optics in Telecommunications}.
\newblock Springer, 2004.

\bibitem{Agrawall1995}
G.P. Agrawall.
\newblock {\em Nonlinear Fiber Optics}.
\newblock Academic Press, 1995.

\bibitem{biochemistry}
P.~Guillamat, C.~Blanch-Mercader, K.~Pernollet, G.~Kruse, and A.~Roux.
\newblock Integer topological defects organize stresses driving tissue morphogenesis.
\newblock {\em Nat. Mater.a}, 21:588–597, 2022.

\bibitem{Yakushevich2004}
L.V. Yakushevich.
\newblock {\em Nonlinear Physics of DNA}.
\newblock Wiley-VCH, 2004.

\bibitem{cosmology}
A.~Vilenkin.
\newblock Cosmic strings and domain walls.
\newblock {\em Physics Reports}, 121(5):263--315, 1985.

\bibitem{Kolb1990}
E.W. Kolb and M.S. Turner.
\newblock {\em The Early Universe}.
\newblock Addison-Wesley, 1990.

\bibitem{Kibble1976}
T.~W.~B. Kibble.
\newblock Topology of cosmic domains and strings.
\newblock {\em Journal of Physics A: Mathematical and General}, 9(8):1387, aug 1976.

\bibitem{Vilenkin1994}
A.~Vilenkin and E.P.S. Shellard.
\newblock {\em Cosmic strings and other topological defects}.
\newblock Cambridge University Press, 1994.

\bibitem{Vachaspati2006}
T.~Vachaspati.
\newblock {\em Kinks and Domain walls: An Introduction to classical and quantum solitons}.
\newblock Cambridge University Press, 2006.

\bibitem{graphene}
A.~Cortijo and M.A.H. Vozmediano.
\newblock Effects of topological defects and local curvature on the electronic properties of planar graphene.
\newblock {\em Nuclear Physics B}, 763(3):293--308, 2007.

\bibitem{Bishop1980}
A.R. Bishop, J.A. Krumhansl, and S.E. Trullinger.
\newblock Solitons in condensed matter: A paradigm.
\newblock {\em Physica D: Nonlinear Phenomena}, 1(1):1--44, 1980.

\bibitem{Eschenfelder1981}
A.H. Eschenfelder.
\newblock {\em Magnetic Bubble Technology}.
\newblock Springer-Verlag, 1981.

\bibitem{Jona1993}
F.~Jona and G.~Shirane.
\newblock {\em Ferroelectric Crystals}.
\newblock Dover, 1993.

\bibitem{Strukov}
B.A. Strukov and A.P. Levanyuk.
\newblock {\em Ferroelectric Phenomena in Crystals: Physical Foundations}.
\newblock Springer-Verlag, 1998.

\bibitem{KUSENKO1997108}
K.~Alexander.
\newblock Solitons in the supersymmetric extensions of the standard model.
\newblock {\em Physics Letters B}, 405(1):108--113, 1997.

\bibitem{AFFLECK1985361}
I.~Affleck and M.~Dine.
\newblock A new mechanism for baryogenesis.
\newblock {\em Nuclear Physics B}, 249(2):361--380, 1985.

\bibitem{KUSENKO199846}
A.~Kusenko and M.~Shaposhnikov.
\newblock Supersymmetric ${Q}$-balls as dark matter.
\newblock {\em Physics Letters B}, 418(1):46--54, 1998.

\bibitem{Wheeler:1955zz}
J.~A. Wheeler.
\newblock {Geons}.
\newblock {\em Phys. Rev.}, 97:511--536, 1955.

\bibitem{Kaup:1968zz}
D.~J. Kaup.
\newblock {Klein-Gordon Geon}.
\newblock {\em Phys. Rev.}, 172:1331--1342, 1968.

\bibitem{Ruffini:1969qy}
R.~Ruffini and S.~Bonazzola.
\newblock {Systems of selfgravitating particles in general relativity and the concept of an equation of state}.
\newblock {\em Phys. Rev.}, 187:1767--1783, 1969.

\bibitem{bazeia2016exact}
D.~Bazeia, M.A. Marques, and R.~Menezes.
\newblock Exact solutions, energy, and charge of stable ${Q}$-balls.
\newblock {\em The European Physical Journal C}, 76:1--13, 2016.

\bibitem{Alonso2023_1}
A.~Alonso-Izquierdo and C.~Garzón Sánchez.
\newblock Families of ${Q}$-balls in a deformed {$O(4)$} linear sigma model.
\newblock {\em Physica D: Nonlinear Phenomena}, 449:133764, 2023.

\bibitem{kolokolov1973stability}
A.A. Kolokolov.
\newblock Stability of the dominant mode of the nonlinear wave equation in a cubic medium.
\newblock {\em Journal of Applied Mechanics and Technical Physics}, 14(3):426--428, 1973.

\bibitem{Alonso2023_2}
A.~Alonso-Izquierdo and C.~Garz\'on S\'anchez.
\newblock Defects composed of kinks and ${Q}$-balls: Analytical solutions and stability.
\newblock {\em Phys. Rev. D}, 107:125004, Jun 2023.

\bibitem{Xie:2021glp}
Q.X. Xie, P.M. Saffin, and S.Y. Zhou.
\newblock {Charge-Swapping Q-balls and Their Lifetimes}.
\newblock {\em JHEP}, 07:062, 2021.

\bibitem{Saffin:2022tub}
P.M. Saffin, Q.X. Xie, and S.Y. Zhou.
\newblock {$Q$-ball Superradiance}.
\newblock {\em Phys. Rev. Lett.}, 131(11):111601, 2023.

\bibitem{ALONSOIZQUIERDO2024}
A.~Alonso-Izquierdo, D.~{Canillas Martínez}, C.~{Garzón Sánchez}, and M.A. {González León}.
\newblock {Geometric construction of non-linear Sigma models with \textit{Q}-ball/\textit{Q}-kink solutions}.
\newblock {\em Chaos, Solitons \& Fractals}, 181:114732, 2024.

\bibitem{Alonso2000}
A.~Alonso-Izquierdo, M.~A.~González León, and J.~Mateos Guilarte.
\newblock Kinks from dynamical systems: domain walls in a deformed {$O(N)$} linear sigma model.
\newblock {\em Nonlinearity}, 13(4):1137, jul 2000.

\bibitem{Alonso2002}
A.~Alonso-Izquierdo, M.~A.~González León, and J.~Mateos Guilarte.
\newblock Stability of kink defects in a deformed {$O(3)$} linear sigma model.
\newblock {\em Nonlinearity}, 15(4):1097, may 2002.

\end{thebibliography}
\end{document}